\documentclass[11pt]{article}
\usepackage{amssymb}

\usepackage{graphics}

\newtheorem{lemma}{Lemma}
\newtheorem{theorem}{Theorem}
\newtheorem{corollary}{Corollary}
\newtheorem{remark}{Remark}
\input{tcilatex}

\begin{document}

\title{Probability around the Quantum Gravity. \\
{\Large Part 1: Planar Pure Gravity}}
\author{V.A.Malyshev \thanks{%
Postal address:INRIA - Domaine de Voluceau, Rocquencourt, BP105 - 78153 - Le
Chesnay Cedex, France.}}

\begin{titlepage}   
 \maketitle 
\begin{abstract}
In  this paper we study  stochastic dynamics which leaves quantum 
gravity equilibrium distribution invariant. We start theoretical study of this dynamics (earlier it 
was only used for Monte-Carlo simulation). Main new results concern the 
existence and properties of local correlation functions in the thermodynamic limit.
The study of dynamics constitutes a third part of the series of
papers where more general class of processes 
were studied  (but it is self-contained), those processes have  
some universal significance
 in probability and 
they cover most concrete processes, also they have many examples in computer 
science and biology. At the same time the paper can serve an introduction to quantum gravity for a probabilist:
 we give a rigorous 
exposition of quantum gravity in the  planar pure gravity case. Mostly 
we use combinatorial techniques, instead of more popular in physics random 
matrix models, the central point is the famous $\alpha =-\frac{7}{2}$ exponent. 
\end{abstract}
\end{titlepage}
\pagebreak
\tableofcontents
\pagebreak
\section{Introduction}

\paragraph{Some history}

I should say some words about the history of discrete gravity. Classical
gravity deals with a smooth (not necessarily four-dimensional) manifold $M$,
pseudo-metric tensor $g_{ij}$ on it and the classical Einstein-Hilbert
action 
\[
S=\int_{M}(\lambda R(x)+L(x)+\mu )\sqrt{\det g}dx 
\]
the various stationary points of which are studied. Here $R(x)$ is the
intrinsic curvature at the point $x$, $L$ - some functional of matter fields 
$\phi (x)$. In the pure (no matter) gravity case $L=0$, we consider only
this case here. Quantum gravity takes into account not only stationary
points but also all other configurations with some weights, that is with a
formal (but which becomes positive for Euclidean metrics) density 
\[
Z^{-1}\exp (-\mu S) 
\]
on some configuration space $\Omega $ of matter fields and metric tensors.
All earlier attempts to do this brought the conclusion that $\Omega $ should
also include smooth structures on $M$ and even $M$ itself, that is the
space, its topology, should be random. Now the only reasonable way to pursue
this program is to discretize everything from the beginning and then to
perform some scaling limits. That is the space becomes a finite complex,
smooth structure becomes a piecewise linear structure, metrics and curvature
are encrypted in one-dimensional and two-dimensional skeletons of the
complex, matter fields are spins which live on the cells of the complex. It
appears that such quantization (discretization) is equally applicable to
other physical systems: relativistic particles, strings etc., but with
different interpretations. For example, the quantized (in such a way) string
consists of a two-dimensional complex (representing a coordinate system and
metrics on the string itself) and spins - vectors in $R^{d}$ which provide a
mapping of the vertices of the complex into the $d$-dimensional Euclidean
space, thus approximating the classical string.

The discretization of the classical gravity was first considered by Regge 
\cite{re} where he gave definitions of some exact mathematical objects
related to the classical general relativity: finite discrete space time, its
curvature and Einstein-Hilbert action. It was afterwards included in the
fundamental monograph \cite{mithwh} but in seventies it was still considered
outside of the main streamline of physics and only rare papers were devoted
to it. Among them however there was a well known paper by S. Hawking \cite
{haw} where the applications to quantum gravity were discussed.

In eighties there are already more than 100 papers concerning discrete
quantum gravity.

In nineties the the number of papers is more than 1000 and still grows at
the moment. Mainly it is due to the appearing algebraic formal techniques to
deal with such problems. This formal techniques follows physical insights on
relations of quantum gravity with string theory, random matrix models etc.
Moreover, recent papers in theoretical physics often contain the following
sententions: ''Two-dimensional random geometry is now placed at the heart of
many models of modern physics, from string theory and two-dimensional
quantum gravity, attempting to describe fundamental interactions, to
membranes and interface fluctuations in various problems of condensed matter
physics'', see \cite{kastwy}.

For a probabilist the quantum gravity is a source of inspiration and also
new mathematics and new philosophy of probability. The paper can serve an
introduction to quantum gravity for a probabilist: it is a mathematical text
on the quantum gravity for the planar pure gravity case.

\paragraph{Dynamics contre equilibrium}

Mostly we consider combinatorial techniques, instead of more popular in
physics random matrix models, the central point is the famous $\alpha =-%
\frac{7}{2}$ exponent. Another goal of the paper is to consider stochastic
dynamics which leaves quantum gravity equilibrium distribution invariant. We
start theoretical study of this dynamics (earlier it was only used for
Monte-Carlo simulation). The study of dynamics constitutes (but mainly it is
self-contained) a third part of the series of papers (see \cite{m1,m2}) where
more general class of processes was studied. These processes have also some
universal character in probability: they cover most concrete processes. Also
they have many examples in computer science and biology.

Here the probability is the classical probability. The quantum gravity
constitutes a bunch (a lot !) of papers overfilling last 10 years well-known
physical journals. Discrete quantum gravity is now considered as a promising
direction towards unifying largest and smallest scales in the nowadays
picture of nature. I consider one part of this field which evidently uses
probabilistic intuition but it is difficult to find even formulations (I do
not mention proofs !) which could be satisfactory for a mathematician: even
when the ''probabilities'' are hopefully positive they are not normalized.
And this is not because of negligence of the authors but because some deep
reasons seem to be behind the curtains.

In the existing physical literature a permanently developing algebraic and
geometric techniques overwhelms the subject. Thus it can be useful to step
away from algebra and geometry, discussing some simple probabilistic aspects
of quantum gravity: even such simple project appeared to rise many natural
but still not answered questions.

There are now two variants in the discrete approaches to quantum gravity:
Quantum Regge Calculus (where links (edges) have lengths as random
variables) and Dynamical Triangulations (where lengths of edges are
constant). The word \textit{dynamical} in the second approach is a little
bit misleading because there is no dynamics at all in this approach: main
techniques uses Gibbs equilibrium distributions on large matrices. That is
why I will call here these approaches equilibrium.

The dynamics appeared earlier in Monte Carlo simulations of quantum gravity.
Here I try to give a probabilistic (not numerical) study of relevant Markov
processes. What is new here (I \ do not know earlier rigorous results) is
that we want to advocate not numerical but analytic and probabilistic
studies of such processes. Why such processes can be useful not only in
computer Monte-Carlo experiments but also as giving theoretical information
? There are many reasons - we give here a short list.

\begin{itemize}
\item  Well known difficulty in averaging over all topologies is that, in 4
dimensions, it includes some questions which are known to be algorithmically
unsolvable. Dynamics substitutes this problem with a new one: instead of
averaging we are looking for a process (with arbitrary initial state) which
will generate all topologies it can generate. This process should have some
symmetries but also it should be a legitimate (for example non-exploding)
stochastic process.

\item  What one would like to have (as in the stochastic quantization in
quantum field theory and Glauber dynamics in statistical mechanics) is a
Markov process leaving Gibbs measure invariant. This is quite natural in
quantum field theory where there are Whiteman axioms and in statistical
mechanics where there is a deterministic dynamics more fundamental than the
Gibbs measure itself. In quantum gravity both these factors are absent and
an alternative viewpoint could be advertised: that the process itself can be
taken to be more fundamental than the Gibbs measure itself.

\item  Dynamics allows to consider the region below the critical point where
equilibrium distribution has no sense. On the contrary this region is even
more natural for the dynamics - like a growing universe (in the computer
time, the term which I know from a paper by A. Migdal). Moreover dynamics
gives also some sense to distributions in the critical point without
performing scaling limits. I do not know the physical counterpart of all
this but its naturalness from probability point of view is evident.

\item  I have absolutely no physical arguments for the choice and even
relevance of the dynamical models, but that is also true for all modern
approaches due to the lack of experimental confirmation. The leading thread
can only be probabilistic intuition and beauty. Relevant question are: what
is universality and generic situation ? It was argued recently, see \cite
{pen}, that computer science could play some role in future physical
theories. Probabilistic aspects which we discuss here make this relation
quite evident by a preliminary model of the universe growing via some
grammar (more exactly a graph grammar) similarly to the random evolution of
a language.

\item  Mathematical thermodynamic theory existing for statistical mechanics
and quantum theory brought many new ideas. It is some surprise that an
attempt to construct similar theory for growing complexes brought quite
unexpected phenomena (see \cite{m1},\cite{m2}) (hopefully having some
physical significance). One of the effects is that one cannot fix an origin
in an infinite universe without Zermelo axiom, any constructive introducing
of a local observer changes drastically the space time in his neighborhood.

\item  It cannot be easy to find critical exponents by Monte-Carlo
simulation because the asymptotic is dominated by the exponential term
which depends strongly on the details of the model. What is usually
simulated is the uniform distribution on the set of triangulations with
fixed number of cells. If we consider a growing complex then we could not
find a Markov process giving the necessary exponents (the famous $\alpha =-%
\frac{7}{2}$) but only some random transformation of measures, that is
called usually a nonlinear Markov processes, giving these exponents.
\end{itemize}

\paragraph{Contents of the paper}

One-dimensional case (section 2) is useful in particular as emphasizing
links between classical probability and two-dimensional quantum gravity.

In section 3 the minimum of necessary definitions are given concerning
complexes and curvature in two dimensional case.

Section 4 contains introductory definitions, problems and some known
results. In section 3.2 we give a short exposition of RMT approach to pure
planar gravity, the only goal of this exposition is to emphasize some
points, related to the combinatorial approach.

In section 5 we study a dynamical model, where the cells are appended in
random to the boundary of the disk. This model is solvable (via random
walks) and we calculate some main quantities. The exponent for this model is 
$-2$ and thus it belongs to a different universality class than models
accepted in physics. But continuum limit in this model is well defined,
gives space with a constant curvature as in the physical model.

Section 6 is the central in the paper. We construct and study nonlinear
Markov processes (where also changes are possible only on the boundary)
which render the equilibrium distribution invariant. We use the Tutte
functional equation method to prove that one gets $-\frac{7}{2}$ exponent.
We develop new combinatorial techniques to study local correlation functions.

In section 7 we consider dynamics where changes can be done elsewhere in the
complex. We study large time behavior of such Markov processes.
\paragraph{Acknowledgments} I thank L. Pastur for elucidating to me some points of the Random Matrix Theory and S. Shlosman for reading the paper and very helpful comments.

\section{One dimensional gravity}

For the physical interpretation of the one-dimensional gravity and many
beautiful calculations we refer to chapter 2 of Ambjorn's lectures \cite{am}%
. Our goal here is to give a probabilistic viewpoint and discuss new
approaches. There is no topology in one dimension: the underlying structure
(cell complex) is one-dimensional - a linear graph.

\begin{figure}[h]
\scalebox{.8} {\includegraphics[0mm,0mm][30mm,20mm]{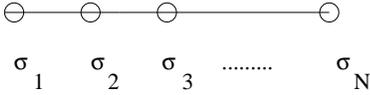}}
\caption{Linear spin graph}
\label{1f5}
\end{figure}
A chain of symbols from some alphabet can be considered as a function on
vertices of such graphs (see Figure \ref{1f5}).

\subsection{Equilibrium distribution}

We give now the basic definition in more abstract terms than in \cite{am},
without prior embedding in Euclidean space: this corresponds more to
Polyakov string quantization. We shall consider distributions on the set of
finite linear spin graphs (sometimes we use the terminology from our
previous papers but mostly it can be skipped). That is the distributions on
the set $\Omega $ of strings (here string comes from computer science
terminology) $\sigma =\left\{ s_{1},...,s_{N}\right\} $, where $N=0,1,2,...$
and $s_{i}\in S$, where $S$ is an alphabet (spin space). For example $S$ can
be the unit sphere in $Z^{d}$ or in $R^{d}$. Case $N=0$ corresponds to the
empty string with no $S$-value prescribed. Define the nonnegative measure on 
$\Omega $ by 
\begin{equation}
Q(\sigma )=\exp (-\mu N-\beta
\sum_{i=0}^{N-1}(f_{1}(s_{i+1})+f_{2}(s_{i},s_{i+1}))  \label{AP}
\end{equation}
for some function $f$. It is convenient to assume that some element $s_{0}$
is fixed. Simplest example is when $s\in Z^{d},f_{2}=0,f_{1}(s)=\infty $ for
all $s$ except a finite set. This measure can be normalized if 
\[
Z=\sum_{\sigma }Q(N,\sigma )<\infty 
\]
Intuitively, a sequence of arrays $%
s_{0},s_{0}+s_{1},...,s_{0}+s_{1}+...+s_{N}$ can be considered as a ''random
walk''. But it is quite different from the classical random walks. We shall
see below its relationship with some computer science problems. We shall
also see how this formal object can be tied to the Euclidean space: in
physical papers one can also see similar steps - abstract object (random
triangulation, internal metrics etc. ) mapped finally to the physical
space-time.

\subparagraph{Simplest examples}

In the first example $S=Z^{d},f_{2}\equiv 0,f_{1}(\sigma )=1$ if $\left|
\sigma \right| =1$ and $\infty $ otherwise, assume $s_{0}=0$. Otherwise
speaking we have the non-normalized distribution $\exp (-l(s))$ on all
possible finite paths $r=(0,s_{1},s_{1}+s_{2},...,s_{1}+s_{2}+...+s_{N})$ in 
$Z^{d}$, starting from $0$ where $N=l(r)$ is the length (number of steps) of 
$r$. It does not always exists. There exists $0<\mu _{cr}<\infty $ such that
the series 
\[
Z=\sum_{s}\exp (-\mu l(r)) 
\]
for the partition function converges for $\mu >\mu _{cr}$ and diverges for $%
\mu \leq \mu _{cr}$. In our case $\mu _{cr}=\ln 2d$.

For the second example $S=R^{d},e^{-f_{1}(s)}=\delta (\left| s\right| -1),$ $%
f_{2}(s,s^{\prime })=\phi ((s,s^{\prime }))$ for some bounded function $\phi 
$ of the angle between the two vectors. For the third example $%
S=R^{d},f_{2}\equiv 0,f_{1}(s)=s^{2}=(s^{1})^{2}+...+(s^{d})^{2}$.

These examples, highly simple and having nothing special from the
probabilistic viewpoint, correspond to one-dimensional analogs of rather
famous actions: free relativistic point particle in $d$-dimensional
space-time ($l$ - length parameter in the Euclidean space, $L=L(0,x)$ is a
path from $0$ to $x$) 
\[
S(L)=\mu \int_{L}dl 
\]
Hilbert-Einstein action ($\kappa $ is the curvature of the curve embedded in
an Euclidean space) 
\[
S(L)=\mu \int_{L}dl+\lambda \int_{L}\left| \kappa \right| dl 
\]
and bosonic string action ($g$ is some metrics on the parameter interval) 
\[
S(L)=\frac{1}{\alpha ^{\prime }}(\int_{L}\sqrt{g(\xi )}d\xi (g(\xi )\frac{%
d^{2}x}{d\xi ^{2}}+\mu ) 
\]
One can consider the introduced distribution as a quantization of the
corresponding classical action. Each link of the discrete path is assumed to
have unit length and thus the length of a path is the number of links.

We shall discuss only the first example, other two are similar, see \cite{am}%
. For $\mu >\mu _{cr}$ we can define the probability distribution on the set
of all finite paths starting from $0$%
\[
P(r)=Z^{-1}\exp (-l(r)) 
\]
Green function are defined as a measure on $Z^{d}$%
\begin{eqnarray*}
G(x) &=&\sum_{s:0\rightarrow x}\exp (-\mu l(r))=\sum_{N}C(N;x)\exp (-\mu N)
\\
&=&\sum_{N}P_{RW}^{(N)}(x)e^{(-\mu +\mu _{cr})N}
\end{eqnarray*}
where $C(N;x)$ is the number of paths from $0$ to $x$ of length $N$, $%
P_{RW}^{(N)}(x)$ - $N$-step transitions probabilities from $0$ to $x$ for
the classical simple random walk in $Z^{d}$. The number of such paths (by
the local limit theorem) is 
\[
C(N;x)\sim (2\pi )^{-\frac{d}{2}}(2d)^{N}N^{-\frac{d}{2}} 
\]
if $x=O(\sqrt{N})$. Green functions have their origin in physics and they
also look like Green functions for a Markov process 
\[
G(0,x)=\sum_{n}p_{0x}^{(n)}=\sum_{L:0\rightarrow x}p(L) 
\]
where $L$ are all paths from $0$ to $x$. But there is no Markov process here.

The following observations are important:

\begin{itemize}
\item  neither $\mu _{cr}$ nor the exponent $-\frac{d}{2}$ do not depend on
the choice of $x$.

\item  $\mu _{cr}$ is not universal, it depends on the dimension and on the
lattice. Also we will get different values of $\mu _{cr}$ if we take
piecewise linear paths in $R^{d}$ with sides of fixed length (see \cite{am});

\item  however the exponent $-\frac{d}{2}$ does not depend on the lattice,
which follows immediately from the local limit theorem;

\item  for $\mu =\mu _{cr}$ the series converges iff $d>2$.
\end{itemize}

Now we want to study the scaling limit $\mu \rightarrow \mu _{cr}+0$. Denote 
$x=s_{1}+...+s_{N}$. In the scaling limit one studies the exponents: mass
(inverse correlation length) exponent $\nu $, susceptibility exponent $%
\gamma $, anomalous dimension $\eta $, Hausdorf dimension $d_{H}$. They are
defined via the leading term behavior for small $\mu -\mu _{cr}$ 
\[
m(\mu )=\lim_{\left| x\right| \rightarrow \infty }\frac{-\ln G(x)}{\left|
x\right| }\approx (\mu -\mu _{cr})^{\nu },\sum_{x}G(x)\approx (\mu -\mu
_{cr})^{-\gamma }, 
\]
\[
G(x)\approx \left| x\right| ^{-d+2-\eta },\sum NC(N;x)\exp (-\mu N)\approx
x^{d_{H}} 
\]

We give now more detailed explanations. We have immediately that 
\[
\chi (\mu )=\sum_{x}G(x)=\sum_{N}e^{(-\mu +\mu _{cr})N}=\frac{1}{1-\exp
(-\mu +\mu _{cr})} 
\]
does not depend on the dimension of $s$ and holds also for more general
spins and interactions (however not always). Thus $\gamma =1$.

If $(\mu -\mu _{cr})x\ll 1$ then $G(x)\approx G_{RW}(x)$, the Green
functions for the simple random walk. That is why $\eta =0$.

Let $p^{(N)}(0,x)$ be the transition probabilities for the simple random
walk on $Z^d$ and $f(p,z)=\sum_N\sum_x$ $p^{(N)}(0,x)e^{ipx}z^N,z\in C,$ -
their generating function. Then 
\[
G(p)=f(p,e^{\mu _{cr}-\mu })=\sum_Ne^{(-\mu +\mu _{cr})N}\left( \frac 1d\sum
\cos p_i\right) ^N=\frac 1{1-e^{-\mu +\mu _{cr}}\frac 1d\sum \cos p_i} 
\]
\[
\approx \frac 1{\mu -\mu _{cr}+\frac 1d\sum \cos p_i} 
\]
which is the classical propagator for quantum relativistic free particle
with mass $\sqrt{\mu -\mu _{cr}}$.

It can be also proved that 
\[
d_H=\nu ^{-1},\gamma =\nu (2-\eta ) 
\]

In the second example different exponents can be obtained for $\lambda
\rightarrow \infty ,\mu \rightarrow \mu _{cr}$.

\subsection{Gravity as a queueing model}

One can construct a reversible dynamics with respect to which the measure $%
Q(\sigma )$ introduced above is invariant. It is a continuous time Markov
chain. The state 
\[
\sigma =s_{1}...s_{N} 
\]
is interpreted as a queue, where $N$ is the length of the queue, $S$ is the
set of customer types and $s_{1}+...+s_{N}$ is a generalized length of the
queue (taking into account customer types and signs of jobs). This is a LIFO
type queue (last in first out) and transitions consist of appending and
deleting links on the right hand side (like arriving and service of
customers in queueing theory). More exactly, for all $\sigma =s_{1}...s_{N}$%
, with rate $\nu $ 
\[
s_{1}...s_{N}\rightarrow s_{1}...s_{N-1} 
\]
and with rate $\lambda $%
\[
s_{1}...s_{N}\rightarrow s_{1}...s_{N}s_{N+1} 
\]
where all values of $s_{N+1}$ are equiprobable. Transitions from the empty
queue $\emptyset $ are $\emptyset \rightarrow s_{1}$ with rate $\lambda $
and equiprobable $s_{1}$.

\begin{lemma}
If $\nu >\lambda $ then the process is ergodic and the distribution (\ref{AP}%
) is invariant with respect to this dynamics.
\end{lemma}

To prove this, note that the restriction $N_{t}=N(\sigma (t))$ of the
process $\sigma (t)$ is also a Markov chain - a birth-death process, its
stationary probabilities are 
\[
Z^{-1}\exp (-\mu N-N\ln 2),\mu =\ln \frac{\nu }{\lambda } 
\]

For models 2 and 3 similar dynamics leaves the distributions invariant. \
Note that a system of two queues would correspond to two interacting
particles etc.

\subparagraph{Supercritical case}

Now we see that they cases $\nu \leq \lambda $ have no sense in the
equilibrium approach but in the dynamical picture they are no worse than the
case $\nu >\lambda $. One cannot write down equilibrium distribution for $%
\nu \leq \lambda $, but at any time moment there exists some distribution $%
Q(t)$ and its limiting properties as $t\rightarrow \infty $ could have
interesting properties. Denote $\sigma (t)$ the string at time $t$.

\begin{theorem}
For $\lambda >\nu $ we have $N(\sigma (t))\rightarrow \infty $ a.s. Moreover
there exist limiting local correlation functions (not too close to the ends
of the string) which define a translation invariant Gibbs field. For example 
\[
\lim_{k\rightarrow \infty }\lim_{t\rightarrow \infty
}P(s_{k}(t)=i)\rightarrow p_{i}
\]
\end{theorem}

In fact for the first example this Gibbs field is a Bernoulli sequence in
all three regions: $\nu >\lambda ,\nu =\lambda ,\nu <\lambda $, see \cite{m3}%
.

\subparagraph{Critical case and scaling limit}

There are two possibilities to consider the critical case.

The first one is to consider the dynamics for critical parameter values. The
properties of this dynamics define the critical exponents. There are results
for sufficiently general transitions: given two positive functions $\nu
(s),\lambda (s,s^{\prime })$, define the transition rates as 
\[
\nu (s_{1}...s_{N}\rightarrow s_{1}...s_{N-1})=\nu (s_{N}),\lambda
(s_{1}...s_{N}\rightarrow s_{1}...s_{N+1})=\lambda (s_{N},s_{N+1}) 
\]
thus depending on the right symbols. Assume that the functions $\nu
(s),\lambda (s,s^{\prime })$ are such that the Markov chain is
null-recurrent, see the conditions in \cite{gaiama}. Let $S$ be finite with
values $a_{1},...,a_{k}$. Let $n(t)=(n_{1}(t),...,n_{k}(t)),n_{i}(t)$ is the
number of symbols $a_{i}$ in the string $\sigma (t)$. Then

\begin{theorem}
The central limit theorem holds for the random vector $n(t)$, that is the
following limit as $t\rightarrow \infty $ exists in distribution 
\[
\frac{n(t)}{\sqrt{t}}\rightarrow \left| w\right| c
\]
where $w$ has the standard gaussian distribution and $c$ is a constant
vector.
\end{theorem}

This gives the same canonical exponents. Note that for the reversible case
the proof reduces to the reflected random walks. The proof \ for
non-reversible dynamics is more involved: for finite $S$ see it in \cite
{gaiama}. For compact $S$ it should be similar. For non-compact it would be
interesting to find examples with non-Gaussian limiting distribution.

The second approach corresponds to the scaling limit in equilibrium case. In
dynamics the parameters are scaled together with time $t$, the parameters
tend to the critical line $\nu =\lambda $ and $\nu -\lambda $ is scaled as $%
t^{-\frac{1}{2}}$. In such dynamics the scaling limit corresponds to the
diffusion approximation in queueing theory. One gets the Brownian motion
with drift for the dynamics of $x$ under the following scaling 
\[
t=\tau N,x=r\sqrt{N},\nu -\lambda =N^{-\frac{1}{2}} 
\]
The drift defines the mass gap in the spectrum of the infinitesimal
generator of the corresponding diffusion process. The proofs here can be
obtained by the application of the techniques known for the critical case.

\subparagraph{Random grammars}

We considered the dynamics, that is called right linear grammar (not
necessarily context free) in the computer science terminology. Now we shall
speak about more general dynamics when transitions can occur at any place of
the string, not only in its right end.

For the first example one can construct the following reversible Markov
chain, leaving invariant the distribution, that appears to be a context-free
random grammar (see \cite{m1}). Each symbol of the string is deleted with
rate $\nu $ and for each $i=0,1,...,n$ we insert a new symbol between
symbols $s_{i}$ and $s_{i+1}$ (where for $i=0$ we put it before $s_{1}$, and
for $i=n$ - after $s_{n}$) of the string $s_{1}...s_{n}$ with rate $\lambda $%
. Appended symbol with probability $\frac{1}{2d}$ will have one of $2d$
coordinate vectors $e$. To prove it note that this this dynamics restricted
to $Z_{+}$, the set of path lengths, is also Markov. \ It is in fact a birth
and death process on $Z_{+}$ with jump rates $q_{i,i+1}=\lambda
(i-1),q_{i,i-1}=\nu i,q_{0,1}=\lambda $. Then its stationary probabilities
are (if $\lambda <\nu $) 
\[
\pi _{k}=\frac{\lambda _{o}\lambda _{1}...\lambda _{n-1}}{\nu _{1}...\nu _{n}%
}\sim C\exp (-\mu k),\mu =\ln \frac{\nu }{\lambda } 
\]
For two other examples the dynamics (not context free) can also be
constructed, we shall do it in another paper in more general cases.

.

\section{Spin Complexes}

\subsection{Cell structures}

Here we present the minimum of basic definitions concerning cell structures.

A complex is obtained by gluing together its elementary constituents -
cells, like the matter consists of molecules. One should be very careful in
defining the rules of gluing and the arising probability distributions. On
the other hand it seems doubtful that some type of cellular structure has
some a priori advantages in front of others. There are no definite physical
reasons to prefer one cell structure or gluing rule etc., over another.
Thus various possibilities should be studied to see what universal laws they
share. In this paper paper we shall encounter two universal classes, one of
them is popular in physics now. Moreover, having some flexibility in
choosing a cell structure one can gain more simplicity in the probabilistic
description and even get solvable models.

\subsubsection{Abstract complexes}

A (labelled) complex $\Gamma $ is a set of elements called cells, there is a
function $\dim A$ on $\Gamma $, the dimension of the cell $A$, taking values 
$0,1,2,...$. The dimension of $\Gamma $ is $\dim \Gamma =\sup_{A}\dim A$.
Let $\Gamma _{d}\subset \Gamma $ be the set of cells of dimension $d$. For
each cell $A\in \Gamma _{d},d>0$, is defined a subset $\partial A\subset
\cup _{i=0}^{d-1}\Gamma _{i}$, the boundary of $A$. Subcomplex $\Gamma
^{\prime }$ of $\Gamma $ is a subset of $\Gamma $ such that if $A\in \Gamma
^{\prime }$ then $\partial A\subset \Gamma ^{\prime }$.

Isomorphism of two complexes is one-to-one mapping respecting dimension and
boundaries. Equivalence classes of complexes with respect to these
isomorphisms are called unlabelled complexes.

The star $St(A)$ of the cell $A$ is the subcomplex containing $A$ and all
cells $B$ such that either $A\in \partial B$ or $B\in \partial A$ or $%
\partial B\cap \partial A\neq \emptyset $.

Note that complexes $\Gamma $ can be considered as particular cases of spin
graphs $(G=G(C),s)$, see\cite{m2}. The correspondence can be constructed in
different ways. For example, let the vertices $i$ of $G$ correspond to cells
of $C$, the function $s(i)$ is the dimension of the corresponding simplex.
Links are defined by the incidence matrix: two vertices $A$ and $A^{\prime }$
of $G$ are connected by a link iff $A^{\prime }\in \partial A$.

Labelled spin complex is a pair $(G,s)$ where $G$ is a complex and $%
s:C(G)\rightarrow S$ is a function on the set $C(G)$ \ of cells of $G$ with
values in some spin space $S$. Isomorphism of two spin complexes is an
isomorphism of the complexes respecting spins. The equivalence classes are
called (unlabelled) spin complexes. Unless otherwise stated we consider only
functions $s$ defined on the cells of maximal dimension; by dualisation it
is often equivalent to functions $s$ restricted to vertices.

\subsubsection{Topological complexes}

There many topological incarnations of abstract complexes. In each of them a
cell is represented by an open disk. A CW-complex is a topological space\ $K$
which is defined by the inductive construction of its $d$-dimensional
skeletons $K_{d}$. Let $K_{0}=\Gamma _{0}$ be a disconnected set of points
(vertices) - cells of dimension $0$. In general, $K_{d}$ is obtained from $%
K_{d-1}$ as follows. Each cell $A$ of dimension $d$ is identified with an
open $d$-dimensional disk $D_{A}$ and some continuous (attaching) map $\phi
_{A}:\partial D_{A}\rightarrow K_{d-1}$ is fixed. Then $K_{d}$ is the factor
space of the union of $K_{d-1}$ and $\cup _{A:\dim A=d}D_{A}$ via
identifications of $x\in D_{A}$ with $\phi _{A}(x)$.

For example, $K_{1}$ is a graph with vertices, zero-dimensional cells, and
links (edges), one-dimensional cells. Link $A$ is a loop if the boundary of $%
A$ is mapped to one vertex. \ Often some restrictions on the attaching maps
are imposed. Here we restrict ourselves to the case $\dim \Gamma =2$ and for
all $A,\dim A=2$, the boundary $\partial D_{A}$ is the union of some cells $%
B,\dim B=0,1$ (in some books, see for example \cite{fofu}, CW-complexes are
defined as already satifying this restriction). With such CW-complex one can
associate an abstract complex with $\partial A=\left\{ B\in K_{0}\cup
K_{1}:B\subset \phi _{A}(D_{A})\right\} $.

We get the class of simplicial complexes (where the cells are called
simplices) if for each 2-cell $A$ its boundary has 3 one-dimensional cells
and the set $\partial A\cap K_{0}$ of vertices uniquely defines $A$. Any
graph without multiple edges, no loops is a simplicial complex.

\subsubsection{Cell surfaces}

In the paper we consider different classes of (two-dimensional) complexes.
The class can be defined either by imposing further restrictions on the
class of complexes defined above or by some constructive procedures to get
all complexes in this class. Anyway such classes are a particular case of a
language defined by some substitutions in a graph grammar, see \cite{m1, m2}.

The following restrictions hold for all complexes in this paper: complex is
a (closed compact) surface. Pseudosurface (closed compact) is a topological
space isomorphic to a finite 2-dimensional simplicial complex with the
following property: each link is contained in the boundary of exactly two
faces (two-dimensional cells). A surface has an additional property that the
neighbourhood of each vertex is homeomorphic to a disk.

This is the list of all compact closed (without holes) 2-dimensional
surfaces. Orientable surfaces are just $S_{\rho },\rho =0,1,2,...,$ - sphere
with $\rho $ handles. Nonorientable surfaces are $P_{1}$ (projective plane), 
$P_{2}$ (Klein bottle), ..., $P_{k},...,$ - sphere in which $k$ holes are
cut and to each of them a Moebius band (crosscup) is attached along its
boundary.

In this case $K_{1}$ is a graph homeomorphically imbedded to the surface $S$%
. Such complexes are studied in the topological graph theory (see \cite{grtu}%
) and in combinatorics, where topological complexes are called maps. Surface
with holes is obtained from a closed surface by cutting out finite number of
disks with non-intersecting boundaries. If the surface has a boundary then
the boundary belongs to $K_{1}$.

Isomorphism of maps is an isomorphism of abstract complexes. In other words,
two maps are called isomorphic if there is a homeomorphism of $S$ such that
vertices map onto vertices, edges on edges, cells on cells.

A map $B$ is a subdivision of the map $A$ if the graph $K_{1}(A)$ of $A$ is
a subgraph of the graph $K_{1}(B)$ of $B$. By Hauptvermutung if two
topological complexes are homeomorphic as topological spaces there exist
their subdivisions isomorphic as abstract complexes.

If the surface is closed the Euler characteristics of the complex is defined
as 
\[
\chi =V-L+F 
\]
where $F$ is the number of faces, $V$- number of vertices, $L$ - number of
links. It does not depend on the complex but only on the surface itself: for
orientable surfaces the Euler characteristics $\chi =2-2\rho $ where $\rho $
is the genus (number of handles), for nonorientable surfaces $\chi =2-n$
where $n$ is the number of crosscups.

We shall use in fact only the following 4 classes.

\paragraph{Arbitrary maps}

This is the class we have just defined. No further restrictions are imposed.
Simplest examples are a vertex inside the sphere (vertex map), an edge with
two vertices inside the sphere - edge map.

\subparagraph{Smooth Cell Surfaces}

Smooth cell surface (see \cite{ste}) is a compact connected smooth
two-dimensional manifold $M$ with finite number of closed subsets (cells) $%
F_{i}$ such that:

\begin{enumerate}
\item  $\cup F_{i}=M$;

\item  for each $i$ there exists a one-to-one smooth mapping $f_{i}$ of $%
F_{i}$ onto a polygon with $n_{i}\geq 3$ faces;

\item  for $i\neq j$ either $F_{i}\cap F_{j}=\emptyset $ or $f_{i}(F_{i}\cap
F_{j})$ is an edge or a vertex of the corresponding polygon.
\end{enumerate}

\paragraph{Triangulations}

This is a smooth cell surface with all $n_{i}=3$. The set $V$ of vertices is
called a cut if there are two subgraphs $G_{1},G_{2}$ such that $G_{1}\cup
G_{2}=G,G_{1}\cap G_{2}=V$. Disk-triangulation is a smooth cell surface,
homeomorphic to the sphere, where there is one distinguished (that will be
the outer face) face and for all other faces $n_{i}=3$, and there is no cuts
with one vertex. Then it can be considered as the triangulation of the disk
(sphere with a hole). For triangulations the absence of cuts is equivalent
to the absence of loops.

\paragraph{Simplicial complexes}

These are triangulations without multiple edges, where moreover every three
edges define not more than one cell. Note that a triangle (cycle of length
3) having inside and outside at least one vertex, is not considered as a
cell.

\paragraph{Convex polyhedra}

Quantizing smooth via piecewise linear structures is possible because the
convex polyhedra have combinatorial counterparts. For example, convex
polyhedra can be considered as maps with $\rho =0$. There is a pure
combinatorial characterization of maps corresponding to convex polyhedra. If
a triangulation has no loops and no multiple edges, then, if $L\geq 4$, it
corresponds (by Steinitz-Rademacher theorem), to a convex polyhedron.

\subsubsection{Local observer (root)}

Labels in complexes are not necessarily given explicitly but the complex is
considered to be labelled if the set $V$ is claimed to be fixed. Labels are
useful for fixing coordinate system in the space but are superfluous for the
geometry and topology. There is a very convenient way to avoid the
superfluous labelling but at the same time giving some algorithmic way to get
a complete coordinatization. Root (local observer) in a (labelled) complex
is an array $(f,l,v)$ where $f$ is a two-dimensional cell, $l$ = its edge, $%
v $ - vertex of $l$. Isomorphism of two complexes with roots
is an isomorphism of complexes respecting the roots. Rooted map (rooted
complex, complex with a local observer) of class $\mathbf{A}$ is an
equivalence class of isomorphisms of complexes with a root in the class $%
\mathbf{A}$ of complexes. Assume that the rooted edge is directed from the
rooted vertex. For disk triangulations we agree that one (the outer) face $i$
is rooted, it is possible that $n_{i}\neq 3$.

\begin{lemma}
The automorphism group of any rooted map is trivial.
\end{lemma}

This is easily proved by induction on the number of cells by subsequent
extending the automorphism from the rooted face to its neighbors.

\subsubsection{Moves}

Graph grammars corresponding to transformations (substitutions here are
called moves) of complexes were studied very little. In the next section we
shall consider Tutte moves, see Fig. \ref{1f4}, which consist in appending an edge
between two vertices of a cell or joining together two disjoint graphs by
identifying two of their vertices. In topology subdivisions played always a
big role. There are two papers (see \cite{al, grva}) where some moves are
studied in detail.

Let $A$ be the commutative associative algebra over $Z_{2}$ (simplicial
chains over $Z_{2}$) generated by the symbols of some (countable) alphabet $%
L $ with commutation relations $s_{i}^{2}=0,s_{i}s_{j}=s_{j}s_{i}$. Thus it
is a linear span generated by the strings (simplices) $\alpha =s_{1}...s_{n}$%
. Define the boundary operator as a linear operator $\partial :A\rightarrow
A $ such that 
\[
\partial \alpha =\sum \beta 
\]
where the sum runs over all subsets of $\alpha $ with the number of elements 
$\left| \alpha \right| -1$. We shall consider here only two dimensional
complexes.

There are other linear operators in this algebra (called Alexander moves) $%
S_{i,j;x},i,j,x\in L$. They are defined as follows. Let $\alpha =ij\beta $,
then 
\[
S_{i,j;x}ij\beta =x(i+j)\beta 
\]

\begin{figure}[tbp]
\scalebox{.8} {\includegraphics[0mm,0mm][70mm,50mm]{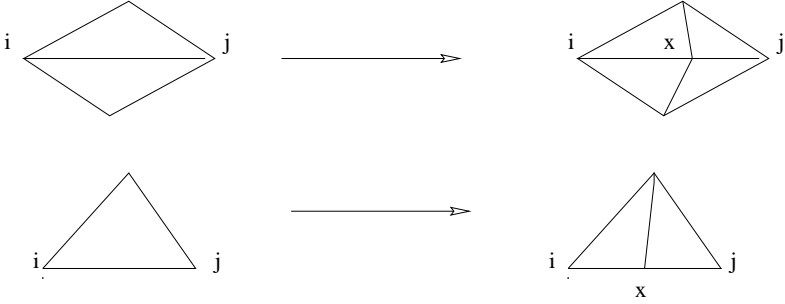}}
\caption{Alexander moves}
\label{1f6}
\end{figure}

Next example: Gross-Varsted moves. 
\begin{figure}[tbp]
\scalebox{.8} {\includegraphics[0mm,0mm][70mm,50mm]{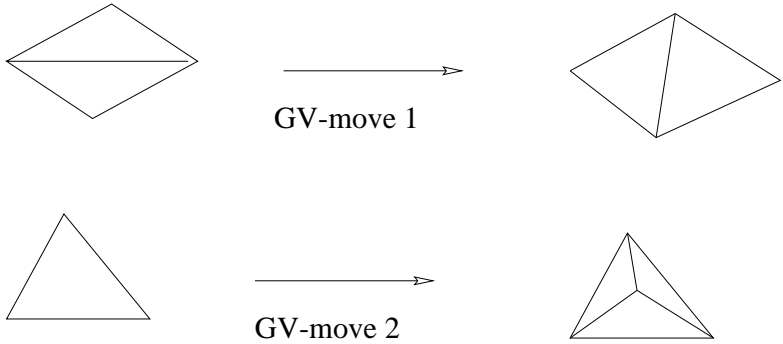}}
\caption{Gross-Varsted moves}
\label{1f61}
\end{figure}
It is proved in \cite{grva} that each Alexandre move can be obtained via
Gross-Varsted moves and vice versa. We say that a set of moves is
irreducible in the class $\mathbf{A}$ of complexes if for each pair $%
T_{1},T_{2}$ of complexes from $\mathbf{A}$ there is a sequence of moves
giving $T_{2}$ from $T_{1}$ (in the physical literature the term ergodic is
used in this cased, but we want to use the standard probabilistic
terminology).

\begin{theorem}
In the class $\mathbf{A}$ of simplicial complexes the set of Alexander moves
and the set of Gross-Varsted moves as well are irreducible.
\end{theorem}

Proof see in \cite{al, grva}.

\subsubsection{Automorphism group}

Let $\mathbf{A}$ be any of the five classes of complexes introduced above.

\begin{theorem}
\label{theAUTO} For most complexes with $N$ two-dimensional cells from $%
\mathbf{A}$ the automorphism group is trivial, that is $\frac{\mathbf{A}%
_{nontrivial}(N)}{\mathbf{A(N)}}\rightarrow 0$ if $N\rightarrow \infty $,
where $\mathbf{A}(N)$ ($\mathbf{A}_{nontrivial}(N)$) is the set of all
complexes with $N$ two-dimensional cells from $\mathbf{A}$ (the same but
with nontrivial automorphism group).
\end{theorem}

Earlier Tutte remarked that it is very intuitive that almost all
triangulations have no nontrivial automorphism. Many rigorous results
appeared afterwards, see \cite{tut6, wor1, wor2}. Proof for the case of
disk-triangulations see in \cite{riwo}.

\subsection{Metrics and Curvature}

The metric structure is defined once it is defined for each closed cell so
that on the edges the lengths are compatible. There are two basic approaches
for defining the metric structure: Dynamical Triangulations - when all edges
have length one and Quantum Regge Calculus - when they are random. We shall
use the first one. Then all cells with the equal number of edges are
identical and on faces the metrics is standard.

One can do it differently. Let first the graph $K_{1}$ be embedded in the
plane, the edges being smooth arcs. Define the metric structure on the graph 
$K_{1}$ so that the edge lengths are all equal to a constant. Inside a cell
with $n$ edges we define the metric structure via some smooth one-to-one
mapping of an equilateral polygon $Q_{n}$ with $n$ edges onto this cell, so
that the smoothness hold also in vicinity of each point on the edge. Then
inside cells the curvature is zero. On edges also: this is shown on the
figure in piecewise linear case.

\begin{figure}[h]
\scalebox{.8} {\includegraphics[0mm,0mm][70mm,50mm]{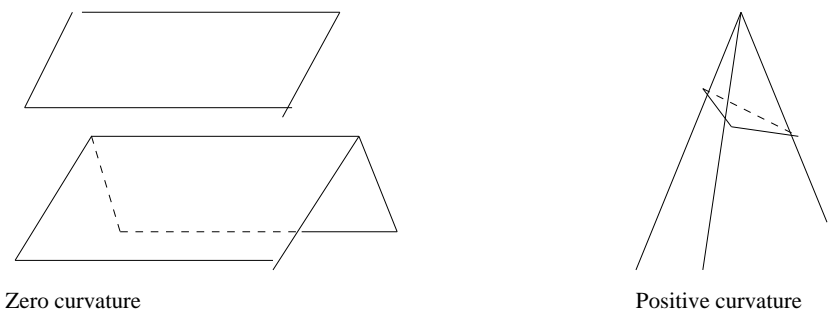}}
\caption{Curvature}
\label{1f1}
\end{figure}

We shall define curvature $R_{v}$ at vertex $v$. As always the curvature is
measured by parallel transport (Levi-Civita connection) of a vector (lying
in the plane in piecewise linear situation) along a closed path: along the
internal part of a triangle as on the Euclidean plane, through an edge - by
unfolding the two half planes separated by this edge to a plane. One sees
immediately that only paths around vertices may give nonzero difference.
Around the vertex $v$ the angle between the initial and the transported
vector is $\varepsilon _{v}=2\pi -\sum_{f}\varphi _{fv}$, where $\varphi
_{fv}$ is the angle of the simplex $f$ at vertex $v$. Note that 
\[
2\pi V-\sum_{v}\varepsilon _{v}=\sum_{v}\sum_{f}\varphi
_{fv}=\sum_{f}\sum_{v}\varphi _{fv}=\pi F 
\]
Using the Euler formula $\chi =V-L+F$ one can get from this the Gauss-Bonnet
formula

\[
\sum_{v}\varepsilon _{v}=2\pi \chi 
\]
for triangulations where $L=\frac{3F}{2}$.

Gauss-Bonnet formula for smooth surfaces is $\int kdA=\frac{1}{2\pi }\int
k(x)\sqrt{g}dx=2\pi \chi $ . Its relationship with the discrete case for a
partition with $n_{i}$-gons with areas $A_{i}$ of the unit sphere (the area $%
A$ of the triangle is $A=\alpha _{1}+\alpha _{2}+\alpha _{3}-\pi $ ) is
given by the formula

\[
\sum k_{i}A_{i}=\sum_{i}\left[ \sum_{j}\alpha _{ij}-(n_{i}-2)\pi \right]
=\sum_{ij}\alpha _{ij}-\pi \sum n_{i}+2\pi F=2\pi V-2\pi E+2\pi F 
\]

Classical examples are: positive curvature - elliptic geometry (sphere,
projective plane); zero curvature, Euclidean geometry (plane, torus, Klein
bottle); negative curvature - hyperbolic geometry (all others).

Now we shall show that the curvature $R_{i}$ at vertex $i$ is defined by the
number $q_{i}$ of edges incident with $i$. Einstein-Hilbert action on the
smooth manifold is 
\[
\int (c_{1}R+c_{2})\sqrt{g}dx 
\]
where $R$ is the Gaussian curvature, $g$ - metrics. It is known that $\int
Rd\sigma =\int R\sqrt{g}dx=4\pi \chi $. Thus the discrete action should be
(up to a constant) $\lambda \rho +\mu N$, where $\rho $ is the genus and $N$
is the number of triangles. We want to write down a discrete analog of this
action with a discrete curvature summing over vertices instead of summing
over triangles. Assume all triangles to be equilateral and scale their area
to 1. Thus each vertex gets area $\frac{1}{3}$ from each incident triangle,
thus $\frac{q_{i}}{3}$ in total. Then $\int Rd\sigma \simeq \sum_{i}\frac{%
q_{i}}{3}R_{i}$ and the formula 
\[
\sum_{i}\frac{q_{i}}{3}R_{i}=4\pi \chi 
\]
holds if only we put the curvature at the vertex $i$ equal $R_{i}=2\pi \frac{%
6-q_{i}}{q_{i}}$.

\section{Equilibrium planar pure gravity}

There are two kind of techniques used in the two-dimensional gravity.
Historically the first one is the combinatorial approach, that was
initialized by Tutte and continued (without any mention of physics) by many
researchers, the papers are published in journals on combinatorics. The
second one is Random Matrix Theory (RMT) approach, that was originated in
physics itself. Calculations in the second approach are very persuasive but
the arguments are not completely rigorous. As far as I know, no explicit
connections between these approaches were established. We use the first
approach and give a short review of the latter approach.

\subsection{Definitions and combinatorial approach}

Let $\mathbf{A}$ be some class of complexes (for example defined in the
previous section) $T$, homeomorphic to the sphere, $\left| T\right| =F(T)$ -
number of cells of dimension $2$ in $T$, $C(N)=\natural \left\{ T\in \mathbf{%
A}:\left| T\right| =N\right\} $. The main example is the class of all
triangulations of a sphere.

The grand canonical ensemble is defined by 
\begin{equation}
P(T)=Z^{-1}\exp (-\mu F(T)),Z=\sum_{T\in \mathbf{A}}\exp (-\mu
F(T))=\sum_{N}C(N)\exp (-\mu N)  \label{ser}
\end{equation}
In particular, the conditional distribution of $T\in \mathbf{A}$ with $N$
fixed is uniform. Easy and general methods to estimate $C(N)$ are useful
sometimes, but can provide only bounds.

\begin{lemma}
(exponential a priori bounds) $c_{1}\gamma _{l}^{N}<C(N)<c_{2}\gamma
_{u}^{N},1<\gamma _{l}<\gamma _{u}<\infty ,c_{i}>0$
\end{lemma}

Proof. Lower bound: this is quite trivial and can be proved in many ways.
For example, take two following complexes homeomorphic to the ring with the
same number of boundary edges from both sides. First one - alternating up
and down triangles (that is standing on an edge and on the vertex
correspondingly), second - two triangles up and two triangles down etc. These
two kind of triangles can be glued sequentially one-after-one in all
possible $2^{n}$ ways.

The following method of proof of upper bounds works even in some more
general situations. One can give an algorithm to construct all possible
complexes with $N$ cells of dimension two. Start with one cell. We enumerate
its edges as $1,2,3$. On each step we add not more than one cell to the
boundary and enumerate new edges immediately after already used numbers. Now
we describe the inductive construction. We take the edge with number one and
make one of the 4 decisions: 1) not to add anymore triangles to this edge,
2) add to it a triangle having exactly two new edges, 3) add triangle to
this edge and to the next edge on the boundary (in clockwise direction), 4)
the same for counterclockwise direction. For each of $4^{k}$ decision
sequences $\omega $ let $f(k,\omega )$ be the number of edges after $k$
steps, $f(k,\omega )\leq 3+2k$. Moreover, if there are $n$ triangles there
cannot be more than $3n$ type 1 decisions.

One needs however exact asymptotics. All known examples exhibit the
following asymptotic behavior 
\begin{equation}
C(N)\sim c_{1}N^{\alpha }c^{N}  \label{asy}
\end{equation}
From (\ref{asy}) it follows

\begin{corollary}
There exists $0<\mu _{cr}<\infty $ such that for $\mu >\mu _{cr}$ the series
(\ref{ser}) converges. It diverges if $\mu <\mu _{cr}$. If $\mu =\mu _{cr}$%
\smallskip\ then $Z<\infty $ iff $\alpha <-1$.
\end{corollary}

Thus, for the parameters $\mu <\mu _{cr}$ the distribution does not exist.
However, the dynamics introduced later allows to consider such $\mu $ and
for them local correlation functions make sense.

No general results are known however. None of the constants $c_{1}>0,c>1,$
is universal, but for all known examples $\alpha $ is. Universality of $%
\alpha $ is not at all simple intuitive fact. For example, predictions based
on physical non-rigorous arguments (see, for example, \cite{zam}) failed to
predict famous $\alpha =-\frac{7}{2}$ in the planar case.

\begin{theorem}
The asymptotics (\ref{asy}) holds for all four classes, defined in the
previous section. Moreover $\alpha =-\frac{7}{2}$ in all cases.
\end{theorem}

Proof. We shall prove it only for triangulations; other cases see in
references cited in ''Enumeration of two-dimensional maps''. In the similar
way we shall define the distribution on the class $\mathbf{A}_{0}$ of rooted
complexes 
\[
P_{0}(T)=Z_{0}^{-1}\exp (-\mu F(T)),Z_{0}=\sum_{T\in A_{0}}\exp (-\mu
F(T))=\sum_{N}C_{0}(N)\exp (-\mu N)
\]
where index zero means that we consider rooted complexes of class $\mathbf{A}
$.

\begin{lemma}
For triangulations 
\[
C(N)\sim (3N)^{-1}C_{0}(N)
\]
\end{lemma}

It follows from triviality of automorphism groups for most complexes (see
theorem \ref{theAUTO}). Then we can take as a root any of $N$ cells of
dimension 2, choose one of its edges and orient it in 2 ways.

Denote $C(N,m)$ the number of disk-triangulations where the outer face has $%
m $ edges, $C_{0}(N,m)$ - where the outer face is moreover rooted. The
following result is similar but can be proved easier.

\begin{lemma}
$C(N,m)\sim m^{-1}C_{0}(N,m)$ for large $N$ and fixed $m$.
\end{lemma}

Proof. Enumerate the edges of the boundary in a cyclic order: $1,2,...,m$.
An automorphism $\phi $ is uniquely defined, if $j=\phi (1)$ is given. We
shall show that almost all complexes do not have an automorphism such that $%
j=\phi (1)$.

To prove this we shall show that for each complex $A$ having a nontrivial
automorphism $\phi $ we can subdivide the complex on two parts $A_{1}\cup
A_{2}=A$ where each cell belongs to only one part, such that $\phi
A_{1}=A_{2}$. This can be done by induction as follows. Take some boundary
edge, take a triangle $T$ with this edge and refer it to $A_{1}$, then put $%
\phi T\in A_{2}$. Each step of induction consists of taking one more
triangle having common edge with already constructed part of $A_{1}$. Now we
can modify $A_{2}$ inside in a number $u(N)$ of ways, bounded from below by
some function $u(N)\rightarrow \infty $ as $N\rightarrow \infty $, uniformly
in $A$. This can be done by choosing $u(N)$ triangles in $A_{2}$, not too
close from each other, and modifying independently some neighborhood of
each keeping the boundary of the neighborhood and the number of cells $n$
in this neighborhood fixed. This is possible as $C(n,b)>1$, where $b$ is
the number of edges on the boundary. Thus for given $A_{1}$ the proportion
of complexes with $\phi A_{1}=A_{2}$ is small.

We have proved that only ''small'' number of complexes have an automorphism $%
\phi $ such that $j=\phi (1)$. As $m$ is fixed then multiplying this number
on $m$ gives again a ''small'' number.

To prove the theorem we should prove that $C_{0}(N)\sim c_{2}N^{-\frac{5}{2}%
}c^{N}$. The universal nature of (\ref{asy}) is strongly supported by the
fact that, for all such examples, the first positive singularity of the
generating function $\sum_{N}C_{0}(N)z^{N}$ is an algebraic singularity,
that gives the asymptotics (\ref{asy}) .

\begin{remark}
\bigskip Assume an algebraic function $y(x)$ is analytic at $0$, has minimal
positive singularity at point $a>0$. We say that its leading exponent is $b$
if there exist such $b_{i}$ and  functions $g_{i}(x)$ analytic at $x=a$ such
that  $y(x)=\sum_{i=1}^{n}g_{i}(x)(a-x)^{b_{i}}+g_{n+1}(x)$. Then we have
the following expansion 
\[
y(x)=\sum_{n}c_{n}x^{n},c_{n}\sim Cn^{-b-1}a^{n}
\]
In our case (for $C_{0}(N)$) $b=\frac{3}{2}$. One could also apply tauberian
theorems in such situation.
\end{remark}

We give some examples where all constants in the asymptotics are known, see
the same references. First example is the class of triangulations defined
above. Here $C_{0}(N)\sim \gamma _{2}N^{-\frac{5}{2}}c^{N},c=3\sqrt{\frac{3}{%
2}}$. For convex polyhedra we have $C_{0}(N)\sim \gamma _{3}N^{-\frac{5}{2}%
}c_{1}^{N},c_{1}=\frac{16}{3\sqrt{3}}$. For simplicial triangulations $%
C_{0}(N)\sim \gamma _{4}N^{-\frac{5}{2}}c^{N},c=\frac{3\sqrt{3}}{2}$. Many
other examples can be given; it is interesting however to understand the
general underlying mechanism.

Tutte \cite{tut1, tut2, tut3} has begun to study the asymptotics for $C(N,m)$
and developed a beautiful and efficient ''quadratic'' method. Afterwards
many authors contributed by developing the method itself and obtaining
asymptotics for various classes $A$ (see review \cite{beri} and more recent
papers \cite{riwo}).

The main idea of Tutte are the following recurrent equations for $%
C(N,m),N=0,1,...;m=2,3,...$ 
\[
C(N,m)=C(N-1,m+1)+%
\sum_{N_{1}+N_{2}=N-1,m_{1}+m_{2}=m+1}C(N_{1},m_{1})C(N_{2},m_{2}),m\geq
3,N\geq 1 
\]
\[
C(0,2)=1,C(0,m)=0,m>2 
\]
These equations are easily derived as follows from the following picture
where the orientation of the rooted edge is marked by arrow, rooted vertex
is the first vertex of the arrow, rooted face is to the right of the arrow
(containing the north pole of the sphere), see Figure \ref{1f4} Take any rooted map
with $(N-1,m+1)$ and do Tutte move 1, take any ordered pair of rooted maps $%
(N_{1},m_{1}),(N_{2},m_{2})$ and perform Tutte move 2. Any rooted map $(N,m)$
can be uniquely obtained in this way. $(0,2)$ corresponds to the so called
edge map with one edge only which is counted twice.

If we introduce the generating function 
\[
U(x,y)=\sum_{N=0}^{\infty }\sum_{m=2}^{\infty }C(N,m)x^{N}y^{m-2} 
\]
the following functional equation 
\[
U(x,y)=U(x,y)xy^{-1}+U^{2}(x,y)xy+1-xy^{-1}U(x,0) 
\]
holds. We shall deduce from this equation that $U(x,0)$ is algebraic and
compute its first singularity, below in this paper, in a bit more general
setting.

\paragraph{Green functions}

Consider a class $\mathbf{A}$ of complexes. Let $\mathbf{A}(m_{1},...,m_{k})$
be a class of complexes, defined with the same restrictions as $\mathbf{A}$,
homeomorphic to the sphere with $k$ holes with $m_{i}$ edges on the
boundaries of these holes. We assume also that these boundaries do not
intersect each other. The Green functions are defined as follows 
\begin{equation}
Z(m_{1},...,m_{k})=\sum_{T\in A(m_{1},...,m_{k})}\exp (-\mu
F(T))=\sum_{N}C(N,m_{1},...,m_{k})\exp (-\mu N)  \label{serGR}
\end{equation}
$Z$ corresponds to the case $k=0$. Rooted Green functions are defined
similarly 
\[
Z_{0}(m_{1},...,m_{k})=\sum_{T\in A_{0}(m_{1},...,m_{k})}\exp (-\mu
F(T))=\sum_{N}C_{0}(N,m_{1},...,m_{k})\exp (-\mu N) 
\]
where the index $0$ everywhere means that we consider complexes with a
distinguished edge on the first boundary with $m_{1}$ edges, the local
observer in the terminology of \cite{m1,m2}. One would like to have an
expression for the Green functions in terms of the basic probabilities (as
for Markov chains).

Green functions are associated with the derivatives $\frac{d^{n}Z}{d\mu ^{n}}%
=-\chi ^{(n)}(\mu )$, that is the factorial moments of $N$.

\begin{lemma}
The partition function and its two first derivatives are finite for $\mu
=\mu _{cr}$ and for $n>2$ we have as $\mu \rightarrow \mu _{cr}+0$ 
\[
\bigskip \frac{d^{n}Z(\mu )}{d\mu ^{n}}\sim c(n)\left( \mu -\mu _{cr}\right)
^{-\alpha -1-n}
\]
\end{lemma}

Proof. We shall see later that $Z$ is an algebraic function of $z=e^{-\mu }$
and has the principal singularity $C(1-\frac{z}{z_{0}})^{-\alpha -1}$ at the
point $z_{0}=e^{-\mu _{cr}}$. In the vicinity of $z_{0}$ we have $1-\frac{z}{%
z_{0}}=1-e^{-(\mu -\mu _{cr)}}\sim \mu -\mu _{cr}$.

This is in a good agreement with the following simple intuitive counting
argument.

\begin{lemma}
For fixed $k,m_{1},...,m_{k}$ there exist constants $0<c_{1}<c_{2}<\infty $
such that 
\[
c_{1}N^{k-1}C_{0}(N)<C_{0}(N,m_{1},...,m_{k})<c_{2}N^{k-1}C_{0}(N)
\]
\end{lemma}

Proof. Take first $k=1$ and prove the upper bound. Take some complex $A\in 
\mathbf{A}_{0}(m)$ with $N$ faces and glue up the hole with some complex $%
B\in \mathbf{A}_{0}(m)$ with $r$ faces where $r$ depends only on $m$. We
shall get some complex $C(A)\in \mathbf{A}_{0}$ with $N+r$ faces. For given $%
A$ and $C\in \mathbf{A}_{0}$ with $N+r$ faces we shall get not more than $p$
complexes $C(A)=C$ where $p$ depends only on $m$. In fact, for any $C$ the
number of subcomplexes with $r$ faces from $\mathbf{A}_{0}(m)$ having the
same root is bounded by $C_{0}(r,m)$. That is why $%
C_{0}(N,m)<cC_{0}(N+r),r=r(m)$. The lower bound can be proved similarly. For 
$k>1$ the proof is similar but one should first choose $k-1$ faces along
which paths with $m_{2},...,m_{k}$ edges will pass. This will give the
factor $N^{k-1}$. This can be done by induction in $k$.

\subsubsection{Uniform asymptotics}

Two questions arise: what is the asymptotics of $C(N,m)$ if both $N,m$ tend
to infinity and what is the asymptotics of other global variables, such as
the number of vertices etc. ? We shall see that these two questions are
related.

$V,L,F$ are well-defined random variables in the grand canonical ensemble
and one could would like to have their joint distribution. In general only
two of them are independent due to the Euler formula $V-L+F=2$. For
triangulations, where each face has 3 incident edges, we have only one
independent variable as $L=\frac{3F}{2}$. For the class of all rooted maps,
where two variables are independent, we have the following lemma.

\begin{lemma}
Let $E(V\mid F=N)$ be the conditional mean number of vertices if the number
of faces is $N$. Then 
\[
E(V\mid F=N)\sim cN
\]
for some $c>0$.
\end{lemma}

As it follows from the formula on p. 157 of \cite{goja} the number of rooted
maps with $N+1$ faces and $m+1$ vertices is 
\[
c(N,m)=\frac{1}{2N-1}C_{2N+m-2}^{m}\frac{1}{2m-1}C_{2m+N-2}^{n} 
\]
Thus 
\[
E(V\mid N)=\frac{\sum_{m}mc(N,m)}{\sum_{m}c(N,m)} 
\]
is defined by the maximum in $m=\alpha N$ of $\ln
(C_{2N+m-2}^{m}C_{2m+N-2}^{n})$ by large deviation asymptotics.

Consider now one-particle Green functions.

\begin{figure}[h]
\scalebox{.8} {\includegraphics[0mm,0mm][90mm,70mm]{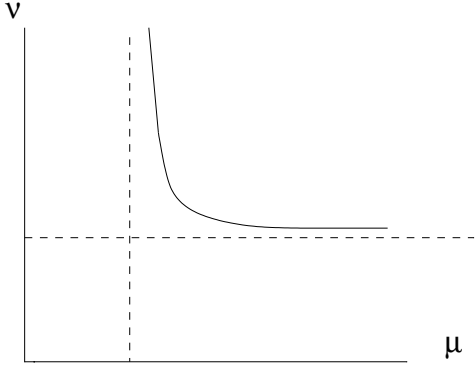}}
\caption{Critical curve}
\label{1f3}
\end{figure}

\begin{lemma}
The following series 
\[
\sum_{N,m}e^{-\mu N-\nu m}C(N,m)
\]
converges above some nondecreasing function $\nu (\mu )$, see Figure \ref
{1f3} Thus the series $\sum_{N,m}e^{-\mu N}C(N,m)$ diverges.
\end{lemma}

Proof. It is quite obvious because the series has all coefficients positive.

Thus we have a family of distributions $P(\mu ,\nu )$. It is of interest to
study the asymptotics and exponents when $\mu =a\nu ,$ where $a,0\leq a\leq
\infty $ is fixed.

The explicit formula (see \cite{goja}) for the number $C_{0}(N,m)$ of
triangulations with a distinguished edge on the boundary (rooted
triangulations) is

\begin{equation}
C_{0}(N,m)=\frac{2^{j+2}(2m+3j-1)!(2m-3)!}{(j+1)!(2m+2j)!((m-2)!)^{2}}
\end{equation}
where $N=m+2j$ is the number of inner cells. $C_{0}(N,m)=0$ if $N-m$ is odd.

Then as $j,m\rightarrow \infty $%
\[
C_{0}(N,m)\sim const\frac{\sqrt{m}}{j(2m+3j)}2^{j+2m}C_{2m+3j}^{j} 
\]
in particular for $j\sim \beta m,\beta >0$%
\[
C_{o}(N,m)\sim const\frac{1}{m^{2}}2^{j+2m-\beta m\log \frac{\beta }{%
2+3\beta }-(2+2\beta )m\log \frac{2+2\beta }{2+3\beta }}\label{jm} 
\]
Thus for all $0<\beta <\infty $ the exponent is $\alpha =-2$. For fixed $m$
the exponent does not depend on $m$ and is $\alpha =-\frac{5}{2}$ and
moreover 
\[
C_{0}(N,m)\sim \phi (m)N^{-\frac{5}{2}}c^{N} 
\]
We have also $\phi (m)\sim Cm^{\frac{1}{2}}c_{1}^{m}$ as $m\rightarrow
\infty $.

\subsection{RMT approach}

Random Matrix Model is the following probability distribution $\mu $ on the
set of selfadjoint $n\times n$-matrices $\phi =(\phi _{ij})$ with the
density 
\[
\frac{d\mu }{d\nu }=Z^{-1}\exp (-tr(\frac{\phi ^{2}}{2h})-tr(V)) 
\]
where $V=\sum a_{k}\phi ^{k}$ is a polynomial of $\phi $ bounded from below, 
$\nu $ is the Lebesgue measure on real $n^{2}$-dimensional space of vectors $%
(\phi _{ii},\func{Re}\phi _{ij},\func{Im}\phi _{ij},i<j)$. It can be written
also as 
\[
\frac{d\mu }{d\mu _{0}}=Z_{0}^{-1}\exp (-tr(V)) 
\]
where $\mu _{0}$ is the Gaussian measure. It is easy to see that $\mu _{0}$
has covariances $\left\langle \phi _{ij},\phi _{kl}^{\ast }\right\rangle
=\left\langle \phi _{ij},\phi _{lk}\right\rangle =h\delta _{ik}\delta _{jl}$%
. Note that for mere existence of the probability measure $\mu $ one needs
that the senior coefficient $a_{p}$ of $V$ were positive and $p$ were even.
In this case there exists a well-developed probability theory of such
models, which we shall not review here, see \cite{pas}.

The fundamental connection (originated from t'Hooft) between RM model and
two-dimensional complexes is provided by the formal series 
\[
\log Z_{0}=\sum_{k=1}^{\infty }\frac{(-1)^{k}}{k!}<tr(V),...,tr(V)>=\sum 
\frac{(-1)^{k}}{k!}\sum_{D_{k}}I(D_{k}) 
\]
where $D_{k}$ is the sum of all connected diagrams with $k$ vertices. Take
for example $V=a_{4}\phi ^{4}$. Then each diagram has labelled vertices $%
1,...,k$, each vertex has labelled thick legs $1,2,3,4$, corresponding to
the product $\phi _{ij}\phi _{jk}\phi _{kl}\phi _{li}$. Each thick leg can
be seen as a narrow strip with two sides, each side is marked with a matrix
index. Dividing by $4^{k}$ we eliminate the numbering of the four legs
leaving them however cyclically ordered. After coupling legs and their sides
(note that coupled sides have the same index and, as each vertex have two
sides with the same index, we get index loops) and summing over indices we
get a factor $n^{N}$ where $N$ is the number of index loops. After this we
are left with 
\[
\sum \frac{(-a_{4})^{k}}{k!}\sum_{D_{k}}h^{L(D_{k})}n^{N(D_{k})} 
\]

For each graph $D$ choose the minimal cell embedding $f(D)$ of $D$ in a
compact orientable surface of genus $\rho $ (topological graph theory \cite
{grtu}). Assume clockwise order of legs. It has $k$ vertices, $2k$ edges and 
$N$ faces. Putting $a=-a_{4}$ and using Euler formula $k=N+2\rho -2$ we have 
\[
\sum \frac{(-a_{4})^{k}}{k!}\sum_{D_{k}}h^{L(D_{k})}n^{N(D_{k})}=\sum_{N,%
\rho }C(N,\rho )a^{k}h^{2k}n^{N}= 
\]
\[
=\sum_{N,\rho }C(N,\rho )(ah^{2}n)^{N}(ah^{2})^{2\rho -2}=a^{2}\exp (-\mu
N-\nu \rho ) 
\]
with $\mu =-\ln (ah^{2}n),\nu =-2\ln (ah^{2})$.

The calculations in RMM can be done only for $n\rightarrow \infty $, thus to
get finite $\mu $ one should scale as $ah^{2}=\frac{b}{n},b=e^{-\mu }$. In
the limit $n\rightarrow \infty $ we have $\nu \rightarrow \infty $ and only $%
\rho =0$ survives giving thus only plane imbeddings. The limit $\lim_{\mu
\rightarrow \mu _{cr}}\lim_{n\rightarrow \infty }$ is called the simple
scaling limit. It was proved (see \cite{beitzu}) that $\alpha =-\frac{7}{2}$
in this case showing again the stability of this exponent.

There are important points in this approach which should be mentioned:

\begin{itemize}
\item  In case $V=a_{4}\phi ^{4}$ the order of all vertices equals $4$, this
is some restriction on the class of maps;

\item  Automorphism group of our labelled diagram factores in two factors.
The first one $C(D)$ related to the permutation of vertices, and second one $%
C_{l}(D)$ related to the permutation of legs in each vertex. Almost all
diagrams have the first factor trivial, but for some of them $C_{v}(D)>1$.
We can sum over nonlabelled diagrams then but each unlabelled diagram will
have a factor 
\[
\sum \frac{C_{l}(D)}{C_{v}D)}
\]
This means that the counting does not coincide with the natural counting
used in the combinatorial approach;

\item  We should fix also $f$ somehow: normally one chooses embedding $f(D)$
to the minimal possible $\rho $. But anyway not all possible triangulations
are taken into account because a given graph can be embedded to surfaces
with different $\rho $. This gives one more reason that the counting rule
does NOT coincide with natural counting where all maps from some fixed class
are counted exactly once. But this should not be taken seriously: anyway
this counting is no worse and no better than others.

\item  There appears a contradiction if one wants to get probability
distributions simultaneously for the matrix model itself and for graph
embeddings. We have probability distribution for the matrix model if $a>0$,
but the probability distribution on the diagrams is achieved only if $a<0$.
Thus one should always perform analytic continuation from $a>0$ to $a<0$.
The free energy for the scaling mentioned above can be rigorously calculated
but the complete argument leading to the graph counting is still lacking.

\item  There are other pure gravity models treated with this approach: more
general pure gravity model counts the number $n(q,T)$ of vertices $v$ with $%
q_{v}=q$:
\end{itemize}

\[
Z=\sum_{T\in A}\prod_{q>2}t_{q}^{n(q,T)} 
\]
where $t_{q}$ are the parameters, see \cite{kastwy}.

\section{Linear boundary dynamics}

The probability distribution 
\[
P(T)=Z^{-1}\exp (-\mu F(T)) 
\]
on some set $\mathbf{A}$ of complexes is invariant with respect to the
following simple Markov process. Let at time $t$ the triangulation be $T(t)$%
. The process is defined by the following infinitesimal transition rates.
With rate $\lambda _{+}(N),N=|T|,$ at time $t$ we destroy $T$, add one more
cell and glue anew all cells randomly together, that is if $N=F(T(t))$ then
we choose uniformly $T(t+0)$ among complexes of the class $A$ with $N+1$
cells. With rate $\lambda _{-}(N)$ we do random choice of a complex with $%
N-1 $ cells.

What dependence on $N$ can be ? If $\lambda _{+}(T)=bf(N),\lambda
_{-}(T)=df(N-1)$ for some positive function $f(N)$ then the probability
distribution $P(T)$ is an invariant distribution with respect to this
process. Proof consists of the remark that the induced process on $N$ is a
reversible Markov chain: a birth and death process on $Z_{+}$ with jump
rates $q_{i,i+1}=bf(i),q_{i,i-1}=df(i-1),q_{0,1}=1$.

The simplest way of Monte-Carlo simulation is to take sufficiently large $N$
and simulate uniform distribution, but it is impossible to find the exponent
in this way. One should compare different $N$ and this can be done via such
a process. Apart from this such dynamics is of no interest, it is not
constructive, especially in higher dimensions. In the rest of this paper we
shall study local dynamics. We start with a simplest local dynamics of
two-dimensional planar complexes. The distribution appears not to be
invariant with respect to the first model dynamics. Thus, there could be two
possibilities: either it will nevertheless give the same exponents for the
invariant measure or its invariant measure belongs to another universality
class (being however irreducible and ergodic). We shall show that the second
one holds.

\subsection{Local Pure Growth}

\begin{figure}[tbp]
\scalebox{.6} {\includegraphics[0mm,0mm][80mm,50mm]{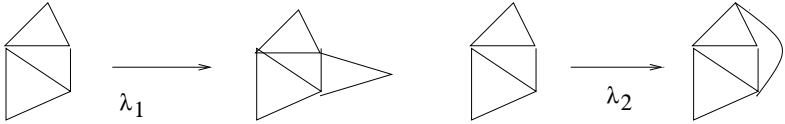}}
\caption{Dynamics on the boundary}
\label{1f2}
\end{figure}

We consider smooth cell surfaces and assume the cells be triangles. One
starts with one triangle and each step consists in attaching a new triangle
on the boundary. There are two kinds of attachment (see Figure \ref{1f2}):
to one or to two edges with the same vertex: for any edge on the boundary we
attach to it a triangle with rate $\lambda _{1}$. For any pair of
neighboring edges on the boundary we attach to them a triangle with rate $%
\lambda _{2}$. At any time the complex is homeomorphic to a closed two
dimensional disk and its boundary - to a circle. We assume that the initial
state is the only triangle and that if the number of edges $m$ on the
boundary is equal to 3 then only $\lambda _{1}$-transitions are possible. We
can consider the states with $m=3$ as giving a triangulation of the sphere
itself (all other states as disk-triangulations), the outside of the
triangle being the cell containing the north pole on the sphere. One can
interpret it as the closing up of the hole in the sphere (the external part
of the complex).

\begin{remark}
It is important to note that one could consider two other variants of this
dynamics. First one is when we consider equivalence classes of cell
surfaces. Then transition rates would be $\lambda _{i}m^{\prime },i=1,2$,
instead of $\lambda _{i}m$, where $m^{\prime }$ is $m$ divided by the number
of automorphisms of the disk triangulation. Second, we shall use its analog
later in more complicated situations, is that there is a distinguished
(rooted) edge on the boundary and transitions can occur only if they touch
this edge.
\end{remark}

There are 3 cases with quite different behavior of this Markov process:
sub-critical or ergodic, critical or null recurrent, supercritical or
non-recurrent. For all these cases we shall study the behavior of local
correlation functions and of the following global variables at time $t$: 
\[
F(t)=F(T(t)),V(t)=V(T(t)),L(t)=L(T(t)),m(t) 
\]
where $L$ is the total number of edges and $m$ is the number of edges on the
boundary.

\subparagraph{Subcritical case}

By definition it is the case when $\lambda _2>\lambda _1$.

\begin{theorem}
If $\lambda _{2}>\lambda _{1}$ then 
\[
\lim_{t\rightarrow \infty }P(m(t)=N)=\pi (m=N)\sim _{N\rightarrow \infty
}CN^{-1}\exp (-\nu N),\nu =\ln \frac{\lambda _{2}}{\lambda _{1}}
\]
Let $\tau _{3}$ be the random number of jumps until first return to the
state $3$ and put $P(T)=P(T(\tau _{3})=T)$ for any triangulation of the
sphere with a distinguished face (outer face). Then
\end{theorem}

\[
P(F(T)=N)\sim C\left( \frac{\lambda _{2}-\lambda _{1}}{\lambda _{1}+\lambda
_{2}}\right) ^{N} 
\]

Proof. Note first that the length $m(t)$ of the boundary is itself a Markov
process: the evolution of the boundary can be seen as the simplest (context
free) random grammar with the alphabet consisting of one symbol $1$
(representing one edge) and with the substitutions

\[
\lambda _1:1\rightarrow 11,\lambda _2:11\rightarrow 1 
\]

This process is obviously reduced to the branching process with one particle
type where $\lambda _{1}$ is the birth rate, $\lambda _{2}$ is the death
rate. Denote this process $m(t)$ - it is a continuous time Markov process
states of which are the points of the lattice interval $[3,\infty )$. It has
jumps $\pm 1$ and the corresponding rates $\lambda _{1}k$ and $\lambda _{2}k$
from the point $k\in \left[ 3,\infty \right] $. Its initial state is $3$.
The process starts anew with the point $3$. The stationary measure for the
process $m(t)$ is 
\[
\pi (m=k)=\frac{\prod_{i=3}^{k-1}\lambda _{1}i}{\prod_{i=4}^{k}\lambda _{2}i}%
\sim Ck^{-1}\exp (-\nu k),\nu =-\ln \frac{\lambda _{1}}{\lambda _{2}} 
\]

\begin{remark}
Note that for dynamics with a local observer the same considerations show
that 
\[
\pi (m=k)=\frac{\prod_{i=3}^{k-1}\lambda _{1}}{\prod_{i=4}^{k}\lambda _{2}}%
\sim C\exp (-\nu k)
\]
The exponents differ, as normal, by $1$.
\end{remark}

Consider the (discrete time) jump process $z_{n}=m(t_{n})$ for $m(t)$, where 
$t_{1}<t_{2}<...$ are the moments of jumps. Let $f(s)=\frac{1}{\lambda
_{1}+\lambda _{2}}(\lambda _{2}+\lambda _{1}s^{2})$ be the generating
function for the transitions of the jump process and $f_{n}(s)$ be its
iterates, let $n(\omega )$ be the first time when $m(t_{n})=3$. It is known
that, see \cite{atne}, 
\[
P(n(\omega )=N)=(f_{N}(0)-f_{N-1}(0))\sim Cm^{N},m=\frac{\lambda
_{2}-\lambda _{1}}{\lambda _{1}+\lambda _{2}} 
\]
This gives the proof.

\subparagraph{Supercritical case}

If $\lambda _{1}>\lambda _{2}$ then the boundary has exponential growth. The
array $\left( V,L,F,m\right) $ behaves like a degenerate branching process
with four particle types and the following rates 
\[
\lambda _{1}m:V\rightarrow V+1,L\rightarrow L+2,F\rightarrow
F+1,m\rightarrow m+1 
\]
\[
\lambda _{2}m:V\rightarrow V,L\rightarrow L+2,F\rightarrow F+1,m\rightarrow
m-1 
\]
Then.for some random variable $\xi $

\[
m(t)\sim \xi \exp (\lambda _{1}-\lambda _{2})t,F(t)\sim c_{1}\xi \exp
(\lambda _{1}-\lambda _{2})t,V(t)\sim c_{2}\xi \exp (\lambda _{1}-\lambda
_{2})t, 
\]
\[
L(t)\sim c_{3}\xi \exp (\lambda _{1}-\lambda _{2})t 
\]
taking into account the Euler formula $V-L+F=1+m$ and $L=\frac{3F}{2}$, the
proof being the same as for similar statements in \cite{m2}.

\subparagraph{Critical case and the exponents}

If $\lambda _{2}=\lambda _{1}$ then the process $m(t)$ is null-recurrent.
One cannot speak about its stationary probabilities but there exists an
infinite stationary measure. It has however the same exponent $\alpha =-1$.

We could consider a different analog of the stationary probabilities: we
consider $P(T(\tau )=N)$, where $T(\tau )$ is the number of cells at the
final moment $\tau $ (when the hole closes up).

\begin{lemma}
Let $\lambda _{1}=\lambda _{2}=\lambda $. Then 
\[
P(T(\tau _{3})=N)\sim \frac{C}{N^{2}}
\]
\end{lemma}

It is well-known, \cite{atne}.

\begin{remark}
The exponent is slightly different from the equilibrium case: $\alpha
=-2\neq -\frac{5}{2}$. But also the following phenomenon occurs. Exponents
in ergodic and null-recurrent case are different: $\alpha =0$ and $\alpha =-2
$ correspondingly. That is the partition function for critical case
converges and the limit of the partition functions as $\mu \rightarrow \mu
_{cr}$ is infinite.
\end{remark}

\paragraph{Local correlation functions}

Even if the exponents are different from the physical theory it would be
interesting to study local correlation functions, they define fluctuations
of the curvature.

Denote by $i$ the $i$-th appeared vertex. Put $q_{i}(t)=0$ if $i$ did not
appear before time $t$ and otherwise put $q_{i}(t)$ equal to the number of
triangles (or edges) incident to $i$. Let $\tau (i)$ be the first time when
the vertex $i$ appeared.

\begin{theorem}
There exist $\chi _{k},\sum \chi _{k}=1,$ such that for $\lambda
_{1}>\lambda _{2}$ 
\[
\lim_{s\rightarrow \infty }\lim_{i\rightarrow \infty }P(q_{i}(\tau
(i)+s)=k)\rightarrow \chi _{k}
\]
For some constants $a,b>0$ and any two vertices $i<j$ such that the initial
distance $\rho (i,j)$ (that is the distance at time $\tau (j)$) between them
equals $d$ 
\[
|P(q_{i}(t)=k,q_{j}(t)=l)-\chi _{k}\chi _{l}|<b\exp (-ad)
\]
for all $i$ and $t$ sufficiently large. If $\lambda _{1}\geq \lambda _{2}$
then 
\[
\lim_{s\rightarrow \infty }\lim_{m\rightarrow \infty }P(q_{i}(\tau
(i)+s|m(\tau (i))=m)=k)\rightarrow \chi _{k}
\]
Also the same exponential decay property holds. Moreover, all local correlation functions $\chi _{k}$ are analytic
functions of $\lambda _{1},\lambda _{2}$ for all values of $\lambda
_{1},\lambda _{2}>0$.
\end{theorem}

\smallskip Proof. For the vertex $i$ put $\xi _{i}(s)=q_{i}(\tau (i)+s)$.
Note that $\xi _{i}(0)=2$ and the number $q_{i}$ will increase until both
adjacent links to this vertex will not enter one new triangle, denote this
random time $\sigma _{i}$, it has exponential distribution with the
parameter $\lambda _{2}$. Thus $\chi _{k}$ is equal to the probability that
the Poisson process with rate $2\lambda _{1}+2\lambda _{2}$ will have
exactly $k-3$ jumps, that is appending triangles to one or two vertices from
the left (or from the right) of $i$. In fact, take vertex $i$ and its two
edges. Until this moment from both sides new triangles in $i$ appear. The
crucial argument is that this process is independent of $m$ as far as during
this time interval $m>3$ and independent on all events which do not touch
vertex $i$.

The second assertion of the theorem is proven quite similarly if we remark
that the processes $q_{i}(t)$ and $q_{j}(t)$ become dependent only when the
distance between $i,j$ becomes less than $4$. For this to occur there should
be many $\lambda _{2}$-events in-between $i$ and $j$, which has
exponentially small probability.

Define the mean curvature 
\[
k=ER=\sum 2\pi \frac{6-q}{q}\chi _{q} 
\]
From the exponential decay it follows the central limit theorem for the
scaled curvature

\begin{theorem}
For any sequence of sets $I$ of vertices 
\[
\frac{\sum_{i\in I}R_{i}-k|I|}{\sqrt{|I|}}
\]
converges to Gaussian distribution as $|I|\rightarrow \infty $.
\end{theorem}

\subsection{Reversible boundary processes}

The reasons why for this dynamics we did not get the desirable invariant
measure on the complexes are rather delicate. We shall give now an intuitive
explanation. One could expect a simple invariant measure for a process which
is reversible. To get a reversible process we should add the possibility of
deletion of faces. For example, let $\lambda _{1}=\lambda _{2}=\lambda $ and
assume that each boundary triangle can be deleted with rate $\mu $ unless
the resulting state does not belong to the class $\mathbf{A}$ of complexes.

\begin{lemma}
This Markov process is reversible.
\end{lemma}

Proof. Let $\alpha _{i}$ be complexes. We shall consider closed paths $%
\alpha _{1},\alpha _{2},...,\alpha _{n},\alpha _{1}$ where each $\alpha
_{i+1}$ is obtained from $\alpha _{i}$ by appending or deleting a triangle
on the boundary. The number of appending in such closed path $\alpha
_{1},\alpha _{2},...,\alpha _{n},\alpha _{1}$ should be equal to the number
of deletions, thus it is $\frac{n}{2}$. Denote $\lambda _{ab}$ the rate of
transitions from complex $\alpha $ to complex $\beta $. Thus 
\[
\lambda _{\alpha _{1}\alpha _{2}}...\lambda _{\alpha _{n}\alpha
_{1}}=\lambda _{\alpha _{1}\alpha _{n}}...\lambda _{\alpha _{2}\alpha
_{1}}=(\lambda \mu )^{\frac{n}{2}} 
\]

There are however complexes with arbitrary $N$ where no triangle can be
deleted, because otherwise we get states, where two graphs intersect only in
one vertex, which are not allowed. The set of such states has a sufficiently
complicated nature and it is difficult to use standard procedure to get
stationary probabilities via balance equations. However, we could split
these graphs into two parts belonging to our class, then we can get
splitting the complex into connected components. It is exactly the latter
operation which brings us to nonlinear Markov processes.

\section{Nonlinear boundary dynamics}

In the preceding section we considered a Markov dynamics, the states were
the complexes themselves. This dynamics was a local dynamics: the changes
could occur at any point of the boundary. We saw that this did not give us
exponents accepted in the physical literature. Here we will construct
dynamics giving exactly the exponent $\alpha =-\frac{7}{2}$. This dynamics
appears not a Markov process. One can think about infinite number of
universes interacting with each other, but not like an infinite particle
system. Simplest pairwise interaction (gluing) of two universes will give
us quadratic functional equations and, as a result, the necessary exponents.

\subsection{Quadratic quasi processes}

Let $\Delta $ be the set of probability measures on some space $S$. We shall
consider a class of transformations $M:\Delta \rightarrow \Delta $,
generalization of Markov chains. These transformations are not generated \
by random maps $S\rightarrow S$, they are nonlinear on $\Delta $.

Let a (denumerable) set $S$ be given and let a Markov chain be defined on
the state space $S$ with transition probabilities $p_{\alpha \beta
},\sum_{\beta :\beta \neq \alpha }p_{\alpha \beta }=1$, from $\alpha $ to $%
\beta $. It defines a linear transformation $L$ 
\[
q=\{q_{\beta }\}\rightarrow Lq=\{\sum_{\alpha }q_{\alpha }p_{\alpha \beta }\}
\]
on $\Delta $. Let also a probability kernel $P((\alpha ,\gamma )\rightarrow
\beta ):S\times S\rightarrow S$ be given, $\sum_{\beta }P((\alpha ,\gamma
)\rightarrow \beta )=1$. It is symmetric $P((\alpha ,\gamma )\rightarrow
\beta )=P((\gamma ,\alpha )\rightarrow \beta )$ and $P((\alpha ,\gamma
)\rightarrow \gamma )=0$. It can be deterministic. It defines the quadratic
transformation $Q$ on $\Delta $ 
\[
q=\{q_{\alpha }\}\rightarrow Qq=\{\sum_{\beta ,\gamma }q_{\beta }q_{\gamma
}P((\beta ,\gamma )\rightarrow \alpha )\}
\]
Then taking a convex combination we have the transformations (formally $%
c_{0}+c_{1}L+c_{2}Q$) on the class of measures on $S$ 
\begin{equation}
q(\alpha ,t+1)=r_{1}\sum_{\beta }q(\beta ,t)p_{\beta \alpha
}+r_{2}\sum_{\beta ,\gamma }q(\beta ,t)q(\gamma ,t)P((\beta ,\gamma
)\rightarrow \alpha )+(1-r_{1}-r_{2})c_{0}(\alpha )  \label{QT}
\end{equation}
for some probability measure $c_{0}(\alpha )$ and nonnegative numbers $%
r_{1},r_{2}$ such that $0\leq r_{1}+r_{2}\leq 1$.We see that the total mass $%
\sum_{\alpha }q(\alpha )=1$ is conserved.

This can be interpreted as follows. Consider a denumerable number of
particles on $S$, $q_{\alpha }$ is the mean number of particles at point $%
\alpha $. With probability $r_{1}$ each particle (independently of the
others) makes a jump according to the probabilities $p_{\alpha \beta }$.
This gives the linear transformation. With probability $r_{2}$ particles
form pairs so that mean number of pairs $(\alpha ,\beta )$ is the product of
means, then each pair $(\alpha ,\gamma )$, independently of each other, gives
birth to one particle at $\beta $ with probability $P(\alpha ,\gamma
\rightarrow \beta )$. Also with probability $1-r_{1}-r_{2}$ one has an
immigration with mean $c_{0}(\alpha )$ of particles to $\alpha $. We want to
emphasize that there is no stochastic process here in its standard sense but
only the transformation of measures. This seems to be related also to field
theory of strings (second quantization of strings) but this physical theory
does not have a mathematical status.

\paragraph{Some probabilistic theory}

There is no theory of such quadratic quasi-processes and we have to discuss
it here. We can rewrite it in more general terms 
\[
T=p_{1}T_{1}+p_{2}T_{2}+(1-p_{1}-p_{2})T_{3} 
\]
Let $k_{1}$ be the contraction coefficient for $T_{1}$, that is for any
probability measures $\mu _{1},\mu _{2}$ we have $\left\| T_{1}(\mu _{1}-\mu
_{2})\right\| \leq k_{1}\left\| \mu _{1}-\mu _{2}\right\| $. Let $k_{2}(x)$
be the contraction coefficient for the stochastic matrix $%
P_{yz}(x)=P((x,y)\rightarrow z)$ and $k_{2}=\sup_{x}k_{2}(x)$.

\begin{theorem}
Assume $p_{1}k_{1}+2p_{2}k_{2}<1$ . If the number of states is finite then
there is exactly one fixed point of $T$ and the convergence to it is
exponentially fast. If the number of states is countable the same assertion
holds under the condition that there exists a (Lyapounov) function $f(x)$
such that $\sum_{x}f(x)<\infty $ and if $\mu (x)\leq f(x)$ then also $(T\mu
)(x)\leq f(x)$.
\end{theorem}

Proof. Take two probability measures $\nu $ and $\mu =\nu +\varepsilon $.
Then we have the following contraction property for $\rho (\nu ,\mu
)=\left\| \varepsilon \right\| $%
\[
\rho (T\mu ,T\nu )=\left\| p_{1}T_{1}\varepsilon +p_{2}\sum \mu
(x)\varepsilon (y)P((x,y)\rightarrow z)+p_{2}\sum \nu (x)\varepsilon
(y)P((x,y)\rightarrow z)\right\| 
\]
\[
\leq (p_{1}k_{1}+2p_{2}k_{2})\left\| \varepsilon \right\| 
\]
Then the first assertion of the theorem follows. To prove the second
assertion note that by compactness there is a fixed point $\nu $ in $%
A=\left\{ \mu :\mu (x)\leq f(x)\right\} $and for each $\mu _{0}$ the
sequence $T^{n}\mu _{0}$ converges to $\nu $.

Continuous time quasi-processes are defined similarly. Instead of
probabilities $p$ we introduce rates $\lambda $: with rate $\lambda _{1}$ we
do the linear transformation, with rate $\lambda _{2}$ the quadratic
transformation, and immigration arrives with rate $\lambda _{0}$. The
equations for the stationary measure are the following 
\[
(\lambda _{1}+\lambda _{2}+\lambda _{0})\pi (\alpha )=\lambda
_{1}\sum_{\beta }\pi (\beta ,t)p_{\beta \alpha }+\lambda _{2}\sum_{\beta
,\gamma }\pi (\beta ,t)\pi (\gamma ,t)P((\beta ,\gamma )\rightarrow \alpha
)+\lambda _{0}c_{0}(\alpha ) 
\]
and can be reduced to the previous case. The time evolution is governed by
the following equation 
\[
\frac{dq(\alpha ,t)}{dt}=\lambda _{1}\sum_{\beta }(q(\beta ,t)-q(\alpha
,t))p_{\beta \alpha }+ 
\]
\[
+\lambda _{2}\sum_{\beta ,\gamma }(q(\beta ,t)q(\gamma ,t)-q(\alpha
,t))P((\beta ,\gamma )\rightarrow \alpha )+\lambda _{0}(c_{0}(\alpha
)-q(\alpha ,t)) 
\]

\subsection{Generating functions}

Now we consider the dynamics with the set $S$ of all plane
disk-triangulations with the root on the boundary and the projection of this
dynamics onto $Z_{+}^{2}$, where the points of $Z_{+}^{2}$ are denoted $%
\alpha =(N,m)$, $N$ is the number of faces (not outer) and $m$ is the number
of boundary edges. That is the measure of the point $(N,m)$ is the sum of
measures of the corresponding complexes. \ This projection will appear to be
also a quadratic quasi-process. Introduce the generating function 
\[
U(x,y)=\sum_{N,m=0}^{\infty }q(N,m)x^{N}y^{m}
\]
We assume homogeneity: for all $\alpha ,\beta ,\alpha _{1},\beta _{1},\gamma 
$ such that $\alpha ,\beta ,\alpha +\gamma ,\beta +\gamma ,\alpha _{1},\beta
_{1},\alpha _{1}+\beta _{1}+\gamma $ all belong to the quarter plane 
\[
p_{\alpha ,\alpha +\gamma }=a_{\gamma },P((\alpha ,\beta )\rightarrow \alpha
+\beta +\gamma )=P((\alpha _{1},\beta _{1})\rightarrow \alpha _{1}+\beta
_{1}+\gamma )=b_{\gamma }
\]
together with the following bounded jumps assumptions: $p_{\alpha \beta }=0$
if $\left| \alpha -\beta \right| >d$ for some fixed integer $d$, $P((\alpha
,\beta )\rightarrow \gamma )=0$ if $\left| \alpha +\beta -\gamma \right| >d$%
, $c_{0}(\alpha )\neq 0$ only for finite number of $\alpha $. Introduce the
generating functions 
\[
A(x,y)=\sum_{n,m}a_{(n,m)}x^{n}y^{m},B(x,y)=%
\sum_{n,m}b_{(n,m)}x^{n}y^{m},C(x,y)=\sum_{n,m}c_{0}(n,m)x^{n}y^{m}
\]
Now we get a functional equation for the generating function of the
stationary measure 
\[
U=r_{1}UA(x,y)+r_{2}U^{2}B(x,y)+(1-r_{1}-r_{2})c_{0}(x,y)+b.c.
\]
where boundary terms can appear because there is no homogeneity in the
vicinity of axes of $Z_{+}^{2}$. If $d=1$ then the boundary terms are linear
combinations of functions $U(x,0)$ and $U(0,y)$ and finite number of $\pi
(N,m)$.

\begin{figure}[tbp]
\scalebox{.8} {\includegraphics[0mm,0mm][120mm,90mm]{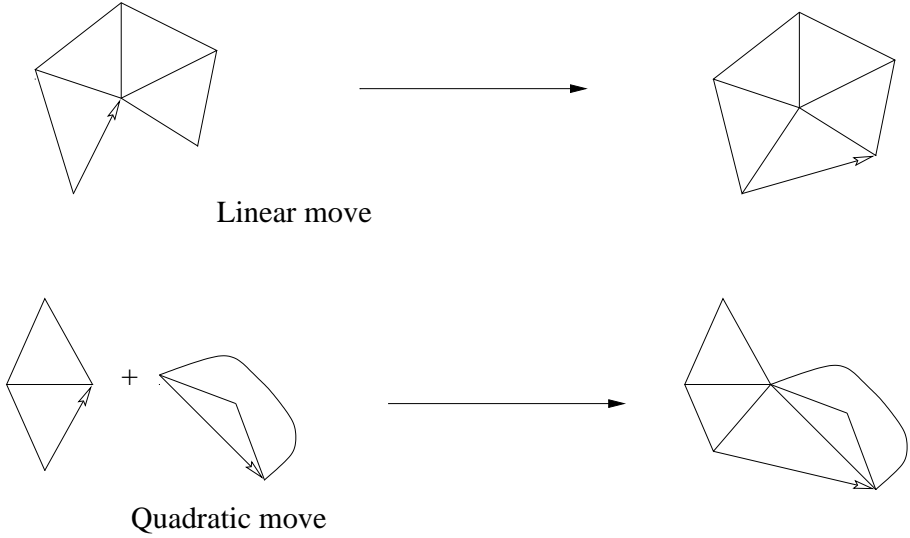}}
\caption{Linear and quadratic moves}
\label{1f4}
\end{figure}

Consider now the case where the jumps correspond to the moves shown on the
Figure \ref{1f4}

Here the jumps are 
\[
p_{(N,m),(N+1,m-1)}=1 
\]
\[
P((N_{1},m_{1}),(N_{2},m_{2}))\rightarrow
(N_{1}+N_{2}+1,m_{1}+m_{2}-1)=1,c_{0}((0,2))=1 
\]
To calculate boundary terms it is convenient to introduce a complex
consisting of one edge with two vertices, it corresponds to $N=0,m=2$. Then $%
c_{0}$ is the unit measure on this complex. Gluing of two such complexes
gives a triangle with $N=1,m=3$. From the line $m=2$ linear jumps are not
possible, this is the only line where the homogeneity is destroyed. Thus the
equation is 
\[
U=r_{1}Uxy^{-1}+r_{2}U^{2}xy^{-1}+(1-r_{1}-r_{2})y^{2}-xy^{-1}r_{1}%
\sum_{N}q(N,2)x^{N}y^{2} 
\]
One can do the following scaling in the main equation 
\[
x\rightarrow xr_{1}^{-1},q(N,m)\rightarrow aq(N,m) 
\]
If we put $ar_{2}=1,\beta =\frac{(1-r_{1}-r_{2})r_{2}}{r_{1}}$ then it
becomes 
\[
U=Uxy^{-1}+U^{2}xy^{-1}+\beta y^{2}-xy^{-1}\sum_{N}q(N,2)x^{N}y^{2} 
\]
We call this equation canonical. Note that the case $\beta =1$ or $r_{1}=%
\frac{r_{2}(1-r_{2})}{1+r_{2}}$ corresponds to the counting problem and was
completely solved by Tutte. We only reproduce his analysis in a more general
setting.

If there exists a solution $U$ with nonnegative coefficients $q(N,m)$ of the
canonical equation then the invariant measure $\widetilde{q}(N,m)$ for the
dynamics will be $\widetilde{q}(N,m)=a^{-1}r_{1}^{N}q(N,m)$.

We shall be interested only in the class $E$ of invariant measures
satisfying the following assumptions:

\begin{enumerate}
\item  (exponential bounds) $q(N,m)<C^{N+m}$ for some $C>0$;

\item  $q(N,m)$ are nonnegative
\end{enumerate}

The main result is the following theorem.

\begin{theorem}
A unique positive invariant measure exists for all $r_{1},r_{2}$. It is
finite iff $\frac{32}{27}\frac{2(1-r_{1}-r_{2})r_{1}r_{2}}{27}<1$\ . The
measure of the sets $\left\{ (N,m)\right\} $ for fixed $m$ is finite iff $%
\frac{2(1-r_{1}-r_{2})r_{1}r_{2}}{27}\leq 1$.
\end{theorem}

In fact we have to prove the following

\begin{lemma}
For any $\beta $ there exists a unique solution of the canonical equation
with positive coefficients in the considered class of measures. The series $%
\sum_{N}q(N,m)<\infty $ converges for each $m$ iff $\beta \leq \frac{2}{27}$%
. The series $\sum_{N,m}q(N,m)$ converges iff $\frac{27}{32}\sqrt{\frac{2}{%
27\beta }}>1$.
\end{lemma}

Proof of the Theorem. One can easily prove that for any initial measure the
difference of measures for  any two complexes with the same $(N,m)$ tends
to zero. Thus, any invariant measure has the property that all complexes
with the same $(N,m)$ have the same invariant measure. Then we can use the
above scaling argument. The theorem follows.

\subsection{Analysis of the functional equation}

Here we prove the Lemma. Putting $U=y^{2}W$ and defining $S_{m}(x)$ by 
\[
W(x,y)=\sum_{m=2}^{\infty }S_{m-2}(x)y^{m-2}
\]
we can rewrite the functional equation 
\begin{equation}
W=\beta +xyW^{2}+xy^{-1}(W-S),S(x)=S_{0}(x)=W(x,0)  \label{feQUA}
\end{equation}
In this case there is only one boundary term $S$ (this simplifies strongly).
This is the generating function for the number of disk-triangulations with
rooted face, having exactly $2$ boundary edges. In the general case the
equations would be more complicated. It demands joining together the
quadratic method of Tutte and the methods developed by the author for
random walks in a quarter plane.

If we know $S_{0}(x)$ as a formal series then all $S_{m}$ are defined
recursively by 
\[
yS_{0}=\beta y+yxS_{1},y^{2}S_{1}=y^{2}xS_{0}^{2}+y^{2}xS_{2},... 
\]
We partially follow the derivation in \cite{goja}. We rewrite the functional
equation (\ref{feQUA}) in the form 
\begin{equation}
(2xU+x-y)^{2}=4x^{2}y^{2}S+(x-y)^{2}-4\beta xy^{3}=D  \label{main}
\end{equation}
Consider the analytic set $\left\{ (x,y):2xU+x-y=0\right\} $ in a small
neighborhood of $x=y=0$. Note that it is not empty, $(0,0)$ belongs to this
set and it defines a function $y(x)=y(x)=x+O(x^{2})$ in the neighborhood of 
$x=0$. We shall prove that $y(x)$ and $S(x)$ are algebraic functions. We
have two equations valid at the points of this set 
\[
D=0,\frac{\partial D}{\partial y}=0 
\]
or 
\begin{equation}
4x^{2}y_{{}}^{2}S(x)+(x-y)^{2}-4\beta xy^{3}=0  \label{pro}
\end{equation}
\[
8x^{2}yS(x)-(x-y)-12\beta xy^{2}=0 
\]
from where we shall get both $y(x)$ and $S(x)$%
\[
x=y(1-2\beta y^{2}),S=\frac{\beta (1-3\beta ^{2}y^{2})}{(1-2\beta y^{2})^{2}}
\]
The algebraic function $y(x)$ satisfies the equation $y^{3}+py+q=0$ with 
\[
p=-\frac{1}{2\beta },q=\frac{x}{2\beta } 
\]
Its discriminant 
\[
\Delta =-4p^{3}-27q^{2}=p^{2}(\frac{2}{\beta }-27x^{2}) 
\]
is not a square in the field of rational functions. Then (see \cite{la}) the
Galois group is $S_{3}$ and the ramification points are $x=\pm \sqrt{\frac{2%
}{27\beta }}$. Cardano (formal) solution of the cubic equation is 
\[
y=\sqrt[3]{-\frac{q}{2}+\sqrt{\frac{q^{2}}{4}+\frac{p^{3}}{27}}}+\sqrt[3]{-%
\frac{q}{2}-\sqrt{\frac{q^{2}}{4}+\frac{p^{3}}{27}}} 
\]
Note that $\left| \frac{q}{2}\right| \neq \left| \sqrt{\frac{q^{2}}{4}+\frac{%
p^{3}}{27}}\right| $ and thus there is no singularities inside the circle of
radius $x_{1}=\sqrt{\frac{2}{27\beta }}$. Two terms with opposite signs give
the cancellation of the lowest order singularity $(x-x_{1})^{\frac{1}{2}}$
and we have thus the leading singularity $(x-x_{1})^{\frac{3}{2}}$. We need
the branch where $y(0)=0$, as $y(x_{1})>0$ then it is known that if the
discriminant is zero then $y(x_{1})=\sqrt[3]{\frac{x_{1}}{2\beta }}$. Then
iterating the equation 
\[
y=\frac{x}{1-2\beta y^{2}} 
\]
we get that the expansion of $y(x)$ at $x=0$ has all coefficients positive. $%
S$ is an algebraic function, analytic for $|x|<\sqrt{\frac{2}{27\beta }}$.
In fact, $S$ could have a pole for $|x|<\sqrt{\frac{2}{27\beta }}$ only if $%
1-2\beta y^{2}(x)=0$ but it would imply $x=0$ which is impossible. To
visualize the expansion of $U$ denote now $y_{0}(x)=y(x)$ and substitute 
\[
S=\frac{\beta (1-3\beta ^{2}y_{0}^{2}(x))}{(1-2\beta y_{0}^{2}(x))^{2}}%
,x=y_{0}(x)(1-2\beta y_{0}^{2}(x)) 
\]
into (\ref{main}). We get 
\[
D=4y^{2}y_{0}^{2}(1-3\beta ^{2}y_{0}^{2})+(y_{0}(x)(1-2\beta
y_{0}^{2}(x))-y)^{2}-4\beta y^{3}y_{0}(x)(1-2\beta y_{0}^{2}(x)) 
\]
\[
=(y-y_{0})^{2}(a(x)+b(x)y) 
\]
as $y=y_{0}(x)$ is a double root of the main equation, and 
\[
a(x)=(1-2\beta y_{0}^{2}(x))^{2},b(x)=-4\beta y_{0}(x)(1-2\beta
y_{0}^{2}(x)) 
\]
Choosing minus sign we have then 
\[
U(x,y)=\frac{y-x}{2x}-\frac{\sqrt{D}}{2x}==\frac{-x+y_{0}(x)}{2x}%
-(y-y_{0}(x))\frac{\sqrt{a(x)+b(x)y}-1}{2x} 
\]
that gives a legitimate expansion.

For given $x$ the convergence radius of $U$ as the function of $y$ is
defined by zeros of $\sqrt{a(x)+b(x)y}$ or $\sqrt{1-\frac{4\beta y_{0}^{2}}{x%
}y}$. As $\frac{y_{0}^{2}}{x}$ increases on the interval $[0,x_{1}]$ then
the convergence radius for $x=x_{1}$ is $R=\frac{x_{1}}{4\beta
y_{0}^{2}(x_{1})}$. We have $R=\frac{27}{32}x_{1}$.

\subsection{Correlation functions}

Using the tree representation introduced below we prove that for most
vertices the conditional distributions of the random variables $q_{v}$
converge to the unique limit as $N\rightarrow \infty $. One should know how
to specify a vertex $v$ if they are not labelled. Below we give some way to
do it, we discussed this problem in \cite{m1, m2} in a wider extent. The
proof is combinatorial but without use of analytic methods, it is
sufficiently involved and we present it not in a completely formal way.

\subsubsection{Tree representation}

\begin{figure}[h]
\scalebox{.8} {\includegraphics[0mm,0mm][50mm,50mm]{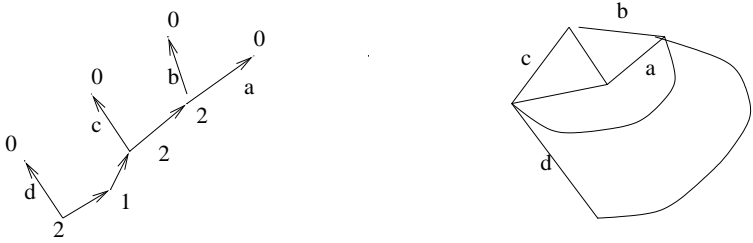}}
\caption{Planar trees and maps}
\label{1f10}
\end{figure}

The generation process of maps, given by recurrent application of Tutte
moves, will be represented as a planar tree. Moreover, this will give a
one-to-one correspondence between maps and some class of planar trees. A
planar tree has a root vertex and grows upwards (it is shown on Figure \ref
{1f10} by arrows). Denote $\mathbf{A}_{0}$ the class of rooted
disk-triangulations. We shall denote vertices of trees by $v$ and vertices
of maps by $w$.

Denote $\mathbf{T}_{0}$ the class of all planar trees characterized as
follows. There can be 3 types of vertices: 0, 1, 2 according to how many
edges go upwards from this vertex. Vertices of type 0 are also called end
vertices. Denote $n_{i}$ the number of vertices of type $i$. Among
2-vertices there are vertices which are incident to one 0-vertex, let their
number be $n_{20}$ and which have two incident 0-vertices, their number is
denoted by $n_{00}$. The only further restriction on this class of trees is
the following. For any vertex $v$ denote $T_{v}$ the tree consisting of the
vertex $v$ and all vertices above $v$. Denote $n_{i}(v)$ etc. - the
corresponding numbers for the tree $T_{v}$. Note that $n_{0}(v)=n_{2}(v)+1$%
. Then the class $\mathbf{T}_{0}$ is characterized by the following
restriction: $n_{0}(v)-n_{1}(v)-1\geq 0$ for all vertices $v$ of type 1.

\begin{lemma}
There is a one-to-one correspondence between $\mathbf{A}_{0}$ and $\mathbf{T}%
_{0}$.
\end{lemma}

Show first that each map generates a planar tree in $\mathbf{T}_{0}$, that
is there is a function $f$ from maps to planar trees. We prove this by
induction on the number of faces. The map itself is represented by the root
vertex of the tree. From the root vertex we draw upwards one edge in case of
move 1 of Tutte, and two edges, in case of move 2, corresponding to
splitting the map on two maps. We can distinguish the latter maps
corresponding to the right or left vertices of the tree accordingly to the
orientation of the rooted edge of the map. In fact, the rooted edges of the
two maps, on which the map is split, have the same orientation. Thus one of
then precedes the other one. The map with the preceding rooted edge we
consider as corresponding to the left vertex (with smaller numbers) and the
map with subsequent rooted edge - to the right vertex. Each step reduces the
number of faces by 1, thus we come to induction hypothesis. If the map is
the edge map (consisting of one edge), then the corresponding vertex is the
end vertex of the tree.

Let us note that $%
N=n_{1}+n_{2},L=n_{0}+n_{1}+n_{2},V=n_{0}+1,m=n_{0}-n_{1}+1 $. Here $N,L,m,V$
refer to the map, and numbers $n_{i},n_{00}$ etc. - to the tree. To prove
the latter equality note that each 1-vertex and each 2-vertex (except those
which are incident to 0-vertices), when passed downwards, diminish $m$ on
one, each of $n_{20}$ vertices gives one more edge to $m$ and each of $%
n_{00} $ vertices gives three more edges to $m$. Thus $%
m=n_{20}+3n_{00}-n_{1}-(n_{2}-n_{00}-n_{20})=n_{0}-n_{1}+1$. As $m\geq 2$
for each map, we have the restriction.

Now show that each planar tree of class $\mathbf{T}_{0}$ generates a map -
we shall show that $f$ is one-to-one. Take a tree and proceed by induction
from upper to lower vertices. All end vertices we declare to be edge maps.
Take $V$ end edges, enumerate them from left to right as the end vertices of
the tree $v=1,2,...,V$. These edges will give $V+1$ vertices in the complex.
We mark all end vertices. Each induction step we take a vertex $v$ such that
there are only marked vertices above it. If $v$ is of type $i$ then the
induction step consists of Tutte move $i$. We mark $v$ after this step. Then
we proceed by induction. Inversely, the map constructed in this way
generates the tree from which we started. All maps are legitimate because $%
m\geq 2$. Lemma is proved.

Contribution of the tree $G(T)$ is defined as the product $%
r_{0}^{n_{0}}r_{1}^{n_{1}}r_{2}^{n_{2}}$ where $n_{i}=n_{i}(T)$ is the
number of $i$-type vertices. This is equal of course to the probability of
the corresponding rooted map.

\begin{remark}
Planar trees are in one-to-one correspondence with the parenthesis systems,
which can be put the product $a_{1}...a_{n}$ in non associative
non commutative algebra, see Figure \ref{1f8}. However, the restrictions
posed on $\mathbf{T}_{0}$ make this one-dimensional grammar more involved. 
\begin{figure}[h]
\scalebox{.8} {\includegraphics[0mm,0mm][50mm,30mm]{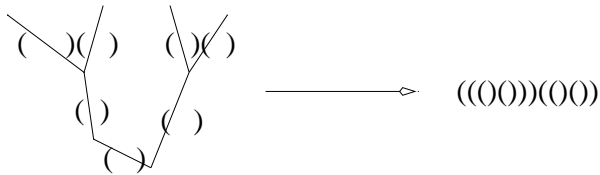}}
\caption{Planar trees and parenthesis}
\label{1f8}
\end{figure}
\end{remark}

\subsubsection{Local curvature}

Take some end vertex (or end edge) $v$ and put $l(v)=\max (v,V-v)$.

\begin{theorem}
If $l(v)\rightarrow \infty $ (necessarily $V\rightarrow \infty $ and $%
N\rightarrow \infty $) then for the equilibrium distribution $P$ there
exists the limit 
\[
\lim_{{}}P(q_{v}=k)=p(k),\sum_{k=2}^{\infty }p(k)=1
\]
\end{theorem}

Proof. \ One can imagine $V$ edge maps, corresponding to the end vertices,
put horizontally along the line and directed from left to right. Take some $%
v $ and the left (that is the rooted) vertex $w=w(v)$ of the corresponding
edge map. We fix orientation of the boundary to be counterclockwise. Define
the history of each vertex $w(v)$. This history can be described in terms of
maps and in terms of the tree, it will be convenient to use both
descriptions. By definition it consists of several parts, that we call
history parts, of the unique path from $v$ to the root of the tree in the
tree $T$ in the downward direction. In the sequence of maps the history ends
either when the vertex $w(v)$ disappears from the boundary of the map (this
means that $q_{w}$ will not change anymore) or in the root of the tree, if $%
w $ is on the boundary of the final map. The first part of the history lasts
until the vertex will be covered by an edge. $q_{w}$ may change only on such
history parts. From the tree point of view the first part of the path starts
with the end vertex $v$, goes downwards vertex by vertex, and lasts until
the first right 2-vertex.We say that the 2-vertex is a right vertex if our
vertex $w(v)$ is in the right map. Next part of the history starts when $%
w(v) $ becomes to belong to the rooted edge of the map and ends exactly as
in the first part.

If the vertex $a$ of the tree is above $v=w(v)$ then $w$ belongs to the
corresponding map $f^{-1}(a)$. Denote $a(w)$ the infimum of $a$ such that $%
a<v$ and $w$ is on the boundary of $f^{-1}(a)$.

Denote $b(w)$ the infimum of $b$ such that $b<v$ and $w$ is still a rooted
vertex in the corresponding complex for each $b_{i}$ on the interval $%
v>b_{1}>b_{2}>...>b$ in the tree. The length $L(v(w))$ of the latter path in
the tree is exactly the difference $q_{w}(j)-q_{w}(k)$ where $j$ is the
first complex of the path (edge map), $k$ - the last one. We shall prove
that $P_{V}(L(v(w))=n)<C\exp (-\gamma n)$ for some $\gamma ,C>0$.  Note
that we can use for this any invariant measure as this conditional
probability is the same for fixed $N,m$ (in fact we are interested only in $%
m=3$).

If from the tree one deletes all 1-vertices then the resulting tree without
1-vertices will be called a bare tree, It defines an equivalence class of
dressed trees, each of them can be obtained by appending some number of
1-vertices to the bare tree. Each nonnegative measure on trees induces a
measure on equivalence classes - bare trees. We start with bare trees.

\paragraph{Case of bare trees}

It is the case when there are no 1-vertices at all, that is $m=n_{0}+1=N$.
In this case all vertices are on the boundary, and once the vertex was
covered it does not participate in the process anymore. It can be covered
only when a left join occurs and until it was covered there can be $k$ right
joins which give $k$ extra edges to $q_{w}$. Any right join joins to $w$ a
complex corresponding to some tree covering an interval to the right of $v$.

For example, let us estimate the probability\ $P(k;j_{1},...,j_{k})$ that
there are exactly $k$ joins to the intervals of lengths $j_{1},...,j_{k}$
covering the interval $I_{r}=\{v+1,...,v+R\}$ before the left join covering
the interval $I_{l}=\{v-L,...,v-1\}$. We shall use the following Markov
property. Call a vertex $v^{\prime }=v(I)$ $I$-separating if it covers
exactly the interval $I$. Then the number of trees with the separating
vertex $v^{\prime }$ is equal to $t(I)t(V-I+v^{\prime })$ where $t(I)$ is
the number of trees on the interval $I$, because $t(V-I+v^{\prime })$ is
also the number of factor trees with respect to the set of trees on $I$.
Thus the probability that $v^{\prime }$ is $I$-separating can be estimated
as 
\[
\frac{t(I)t(V-I+v^{\prime })}{t(V)}\sim \min (b,cI^{-\frac{3}{2}}) 
\]
as $t(n+1)=\frac{1}{n+1}C_{2n}^{n}\sim cn^{-\frac{3}{2}}2^{n}$ are
well-known Catalan numbers and $b<1$. It follows that the probability $P_{v}$
that the vertex $v$ will have a right join earlier than a left join. It is
clear that $P_{v}<d<1$. Then using the Markov property we shall get by
induction for large $r$ and some $a<1$%
\[
P(q_{v}\geq r)<a^{r} 
\]
Thus we got the exponential estimates. The existence of the thermodynamic
limit follows from the fact that the influence of the boundaries takes place
also with probabilities less than $a^{l(v)}$.

\paragraph{Case $n_{1}>0$.}

We fix a bare tree and consider one auxiliary problem (urn problem)
concerning the distribution of 1-vertices on the bare tree. The estimates
are uniform in bare trees (equivalence classes).

Consider first the probability that the vertex $w$ will get large value of $%
q_{w}$ due to 1-vertices until it will covered at the first time. We shall
do such estimates separately for each history part and consider in detail
only the first history part. Let $v$ be the vertex where the first parts
ends. Let $T_{v}$ be the tree over this vertex (with the root $v$) and $%
f^{-1}(v)$ be the corresponding complex.

First consider the case when $T_{v}$ has only one 00-vertex, then 0-vertices
join sequentially to already existing trees. It means that different ways to
put 1-vertices to the bare tree can be identified with all possible ordered
arrays of nonnegative integers $a_{1},...,a_{n}$ such that for all $%
k=1,2,...,n$ we have $\sum_{i=1}^{k}a_{i}\leq k-2$.

We can formulate the following abstract urn model. Let we have $n$ urns and $%
m$ balls in these urns, $a_{i}$ - the number of balls in the urn $i$. Let $%
c(n,m)$ be the number of arrays $a_{1},...,a_{n}$ such that $%
\sum_{i=1}^{k}a_{i}\leq k,\sum_{i=1}^{m}a_{i}=m\leq n$. Then we have the
following recurrence 
\[
c(n,m)=\sum_{k=0}^{m}c(n-1,m-k) 
\]
or 
\[
c(n,m)=c(n,m-1)+c(n-1,m-1) 
\]
from where it is not difficult to get asymptotics for the number $%
c(n,m;i)=c(n-1,m-i)$ of arrays among $c(n,m)$ such that there are exactly $i$
balls in the last urn. Then we have an explicit formula for the generating
function 
\[
f(x,y)=\sum_{n=-\infty }^{\infty }\sum_{m=1}^{\infty }c(n,m)x^{n}y^{m}=\frac{%
y}{1-y(1+x)^{{}}}\sum_{k=1}^{\infty }kx^{k} 
\]
the coefficients coincide with ours for $n\geq m$. \ We want to prove that $%
\frac{c(n,m)}{c(n,m-1)}<b$ for some $b<1$, and to find a method (not using
generating functions) which could work in the general situation. For this we
rewrite $c(n,m)$ in terms of the number of paths starting at the line $m=1$
and ending at the point $(n,m)$. Steps of the paths are either $(0,1)$\ or $%
(1,1)$. We have 
\[
c(n,m)=\sum_{k=m+1}^{n}kL(k;n,m) 
\]
where $L(k;n,m)$ - the number of paths from the point $(k,1)$ to the point $%
(n,m)$, as $c(k,1)=k$. As $L(k;n,m-1)=L(k-1;n-1,m-1)$, then $\frac{c(n,m-1)}{%
c(n-1,m-1)}=1+O(\frac{1}{m})$. The result follows.

Consider now the general case. Instead of the urn problem on the interval $%
[1,n]$ we have ah urn problem on an arbitrary planar tree $T$ under the
conditions 
\[
\sum_{i\in T_{v}}^{{}}a_{v}\leq V(T_{v}),\sum_{v\in T}^{{}}a_{v}=m 
\]
where $a_{v}$ is the number of balls in the urn (vertex) $v$ of the tree, $%
V(T_{v})$ is the number of vertices of the tree $T_{v}$. Let $c(T_{v},m)$ be
the number of such arrays on the tree $T_{v}$ with $m$ balls. If for example
from the vertex $v$ only two edges go upwards to the vertices $v(1),v(2)$,
then 
\[
c(T_{v},m)=\sum_{i=0}^{m}c(T_{v},m;i),c(T_{v},m;i)=%
\sum_{m_{1}+m_{2}=m-i}c(T_{v(1)},m_{1})c(T_{v(2)},m_{2}) 
\]
Then the argument is similar to the previous one. Note that $%
c(T_{v},1)=V(T_{v})$. We want to compare $c(T_{v},m;i)$ and $c(T_{v},m;i+1)$%
, for this we iterate the latter recurrent equation for $c(T_{v},m;i+1)$ to
the very end, that is we get the sum of terms $B_{s}^{i+1}$, in each of them
all factors equal $c(T_{v^{\prime }},1)$ for some $v^{\prime }$. The
iteration process for $c(T_{v},m;i+1)$ there corresponds the similar process
for $c(T_{v},m;i)$, that is why to each term $B_{s}^{i+1}$ there corresponds
the term $B_{s}^{i}$ in the expansion for $c(T_{v},m;i).$ In that term one of
the factors is $c(T_{v^{\prime }},2)$ instead of the factor $c(T_{v^{\prime
}},1)$ in the term $B_{s}^{i+1}$. Thus as before $\frac{c(T_{v},m)}{%
c(T_{v},m-1)}\gtrsim 2$. From this bounds uniform in bare trees follow. The
influence of the boundary is exponentially small.

Similarly one can estimate other correlation functions, for example, the
decay of correlations.

\begin{theorem}
Let $V\rightarrow \infty $ and take two vertices $v_{1},v_{2}$ with $%
l(v_{i})\rightarrow \infty $. Then 
\[
\left| \left\langle q_{v_{1}}q_{v_{2}}\right\rangle -\left\langle
q_{v_{1}}\right\rangle \left\langle q_{v_{2}}\right\rangle \right| <c\exp
(-\alpha \left| v_{1}-v_{2}\right| )
\]
\end{theorem}

\section{Internal dynamics}

We considered above only a growth of the boundary, that was quite natural:
many modern technologies follow this principle. But also another dynamics is
possible where all cells (even inside the building) can evolve. We shall
consider here some questions related to such dynamics.

Note that Gross-Varsted moves can be used not only for simplicial complexes
but for other classes as well, as it is seen from the picture. Consider
GV-moves 1 and 2 and the inverse one to 2, consider the Markov chain with
rates $\lambda _{i},i=1,2,\mu $ for these moves correspondingly.

\paragraph{Thermodynamic limit of local processes}

If $\lambda _{2}=\mu =0$ then $V,L,N$ are invariants. Let $A^{\prime
}\subset A(N,L)$ be an irreducible component of the set of (nonequivalent)
complexes with given $N$ and $L$ and $C(A^{\prime })=|A^{\prime }|$. We make
an assumption that a move can only be done if it gives non-equivalent
complex. We formulate the following lemma without proof.

\begin{lemma}
If $\lambda _{2}=\mu =0$ then the Markov chain on each $A^{\prime }$ is
reversible with respect to the uniform measure. On the class of simplicial
complexes this component coincides with the whole class.
\end{lemma}

Proof. Reversibility is verified via the condition $\pi _{\alpha }\lambda
_{\alpha \beta }=\frac{\lambda }{C(A^{\prime })}=\pi _{\beta }\lambda
_{\beta \alpha }$ if $\lambda _{\beta \alpha }=\lambda _{\alpha \beta
}=\lambda $.

The following example shows that large time and large $N$ limits are not
interchangeable, that is 
\[
\lim_{t\rightarrow \infty }\lim_{N\rightarrow \infty }\neq
\lim_{N\rightarrow \infty }\lim_{t\rightarrow \infty } 
\]
for local quantities. This the simulation is slow and dangerous in this
case. Consider the sequence of such chains $\xi ^{(N)}(t)$ having the
embedded state spaces 
\[
...\subset A^{N}\subset A^{N+1}\subset ... 
\]
Take a vertex $v$ at time $0$ and consider random variables $q_{v}^{(N)}(t)$
- number of edges at $v$ at time $t$. We have $L=\frac{3N}{2},V=\frac{N+4}{2}
$ and it could be natural to think that $q_{v}\approx \frac{L}{V}\rightarrow
3$. But the following argument shows more complicated situation.

\begin{lemma}
Consider the class of simplicial complexes. As $N\rightarrow \infty $ the
limiting process exists and is the random walk on $[3,\infty )$ with
transition rates $\lambda _{i,i+1}=\lambda _{i,i-1}=\lambda i$. Thus $%
P(q_{v}^{(N)}(t)=k)\rightarrow _{N\rightarrow \infty }0$.
\end{lemma}

Proof. For fixed $N$ the process $q_{v}^{(N)}(t)$ is Markov with state space 
$3,...,N$ with rates $\lambda _{i,i+1}=\lambda _{i,i-1}=\lambda i$. In fact,
each edge incident to $v$ can be changed to a transversal and, for each
triangle containing $v$, its edge not containing $v$ can be erased by
GV-move, this will give one more incident edge. The limiting random walk is
null recurrent and thus big fluctuations in it occur until it reaches
equilibrium for fixed $N$.

Similar proof does not hold for other classes of complexes.

Now consider Markov chains where the only transitions are A-moves. \ To get
ergodic chains we change the generator which produces jumps. Now the jumps
are produced by any vertex $i$ with rates $\lambda $ or $\mu $. For fixed $i$
with rate $\lambda $ take randomly (that is with probability $q_{i}^{-1}$)
one of the edges on the boundary of $St(i)$ and do the A-move corresponding
to this edge. Let $\mu $ be the rate of the inverse A-move at vertex $i$,
also for each possible vertex $v$ of degree $4$ on $\partial St(i)$ with
equal probability we take one pair of triangles (on the right hand side of
the A-move) and do the inverse A-move. Once the vertex $i$ appeared it can
disappear afterwards. Let $t(i)$ the time when vertex $i$ appeared.

\begin{theorem}
If $\lambda >\mu $ then $q_{i}(t)\rightarrow \infty $ with positive
probability. If $\lambda <\mu $ then the vertex disappears a.s. and $%
Eq_{i}(t)$ is uniformly bounded.
\end{theorem}

Proof. Let for each vertex $v$ $a(t)=a_{v}(t)$ be number of vertices $j$ on $%
\partial St(v)$ with $q_{j}=4$, let $b(t)=q_{v}(t)-a(t)$. Fix vertex $i$. If 
$v\in \partial St(i)$ then $v+1$ is the next vertex on $\partial St(i)$ in
the clockwise direction.

Consider the process $(a_{i}(t),b_{i}(t))$ and for fixed configuration
outside $St(i)$ write down its infinitesimal jumps in $Z_{+}^{2}$. For a
direct and inverse A-move there can be only three possibilities:

\begin{figure}[h]
\scalebox{.8} {\includegraphics[0mm,0mm][90mm,100mm]{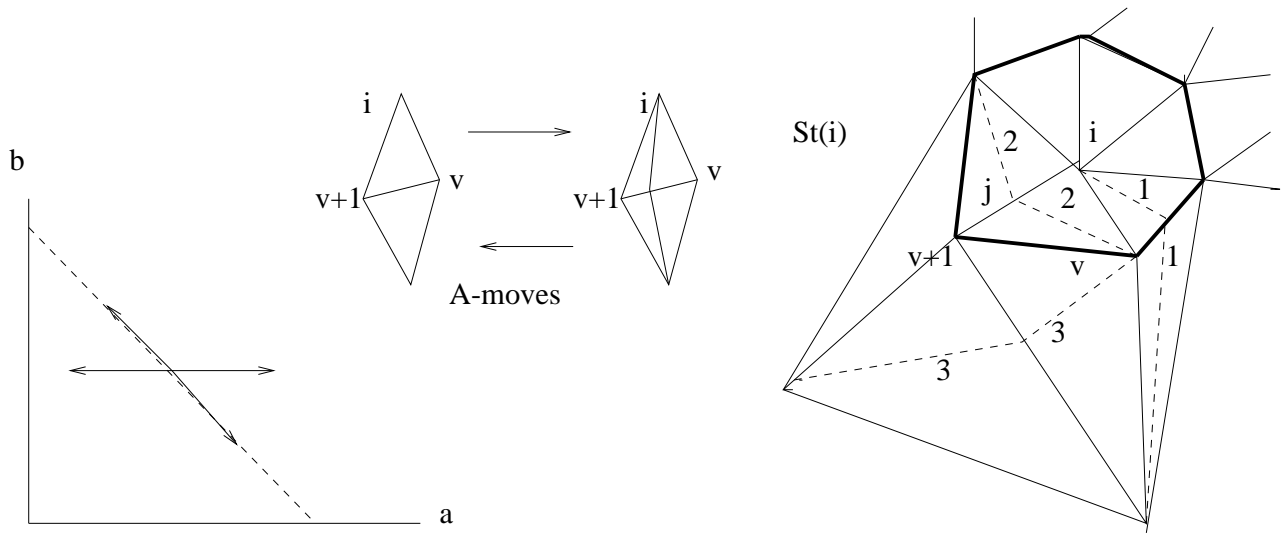}}
\caption{Proof of the theorem}
\label{1f7}
\end{figure}

\begin{enumerate}
\item  Two edges (marked 1 on the figure) appear on some link ($(v-1,v)$ on
the figure). Thus here the transition is $a,b\rightarrow a+1,b$ with rate $%
2\lambda \frac{q}{q}=2\lambda $. Here and further factor $2$ because the
same move can be produced also by the opposite vertex. Inverse move $%
a,b\rightarrow a,b+1$ with rate $2\mu \frac{a}{a}=2\mu $;

\item  This move is produced by vertex $v\in \partial St(i)$ (dotted edges 2
on the figure), the new vertex appears on the edge $(i,v+1)$. It produces a
change in the vector $(a,b)$ only if $q_{v+1}\neq 4$. Thus here $%
a,b\rightarrow a+1,b-1$ with rate $2\lambda a\frac{b}{q_{v}}$. Inverse move
gives the jump $a,b\rightarrow a-1,b$ with rate 2$\alpha \mu \frac{a}{a_{v}}$%
.

\item  Next move is also produced by vertex $v$ (edges 3 on the figure). $%
q_{v}$ can be transformed $4\rightarrow 3,5\rightarrow 4$.
\end{enumerate}

In fact we do not need rates for 2 and 3: note only that these jumps
conserve $q_{i}=a+b$. Assume first $\mu >\lambda $. Then the embedded
process $f(n)=q_{i}(t_{n})=a(t_{n})+b(t_{n})$, where $t_{n}$ are the jump
moments, satisfies the following inequality 
\[
E(f(n+1)\mid f(n))<f(n)-\varepsilon 
\]
for some fixed $\varepsilon >0$. By the submartingale techniques (see, for
example, \cite{famame}) we have the proof. In the opposite case we have 
\[
E(f(n+1)\mid f(n))>f(n)+\varepsilon 
\]
and again the techniques of \cite{famame} works.

It seems plausible that if $\lambda <\mu $ then for all sequences $%
t(i)\rightarrow \infty $ the process $q_{i}(t)$ tends to some proper
distribution if $s=t-t(i)$ is fixed. If $\lambda \ll \mu $ it can be proved.
On the contrary for the critical case $\mu =\lambda $ random variables $%
q_{i}(t)$ fluctuates as for the Brownian motion. Compared with the results
in the previous section this gives argument that we do not get the physical
invariant measure here.

\end{document}